%% file: main.tex
\documentclass[11pt, oneside]{article}
\usepackage[font=small,labelfont=bf]{caption}

\input{macros_wenjia_summary.tex} 
\usepackage{natbib}
 \bibpunct[, ]{(}{)}{,}{a}{}{,}%
 \def\bibfont{\small}%

\title{Sales Policies for a Virtual Assistant}
\author{Wenjia Ba, Haim Mendelson and Mingxi Zhu\footnote{To contact the authors, email wenjiaba, haim, mingxiz@stanford.edu.}}
\affil{Graduate School of Business, Stanford University, Stanford, CA 94305}

\date{August 2020}
\begin{document}

\maketitle
\input{sec0_abstract.tex}

% \tableofcontents
% \newpage
\input{sec0_Intro.tex}

\input{sec1_Literature.tex}

\input{sec1_Overview.tex}
\input{sec2_ModelExp.tex}

\input{sec3_EquilirbiumPrice.tex}
\input{sec4_OptimalRanking.tex}

\input{sec5_Implications.tex}

\input{sec5_webInterfaceComparison.tex}
\input{sec7_conclusion.tex}

%  Bibliography
\newpage
\bibliographystyle{informs2014}
\bibliography{bibVA}

% Appendices
\newpage
\begin{center}
\huge \textbf{Appendices}
\end{center}
\appendix
\input{app1_notation.tex}
\input{app2_proofs.tex}

\end{document}

%% file: macros_wenjia_summary.tex
\usepackage{sectsty}
\usepackage[utf8]{inputenc}
\usepackage[english]{babel}
\sectionfont{\fontsize{15}{18}\selectfont}
\newtheorem{theorem}{Theorem}[section]

\newtheorem{proposition}[theorem]{Proposition}

\newtheorem{algrm}[theorem]{Algorithm.}

\usepackage{subfiles}
\usepackage{import}
\usepackage{hyperref}
\usepackage{amssymb}

\usepackage{amsbsy}
\usepackage{amsmath}
\usepackage{bm}
\usepackage{color}
\usepackage{multirow}
\usepackage{booktabs}
\usepackage{wasysym}
\usepackage{lmodern} % to avoid font size not available warning
\usepackage{textcomp} % to avoid font warning for textbullet used in \item
\usepackage{wasysym} % useful symbols
\usepackage{dashbox} % for use of dashd box \dbox{...}
\usepackage{mathtools} % for \DeclarePairedDelimiter below
\usepackage{tabto}  % for \NumTabs{3} and \tab commands
\usepackage[ruled,vlined]{algorithm2e}
\usepackage{enumitem}
\usepackage[pdftex]{graphicx}
\usepackage{subfig}

\usepackage[margin=1in]{geometry}  
\usepackage{float}
\usepackage{booktabs}
\usepackage{bbm}
\usepackage{caption}
\usepackage{algpseudocode}
\usepackage{setspace}
\usepackage{varwidth}
\usepackage{multirow}
\usepackage{rotating}
\usepackage{array}
\usepackage{tikz}
\usetikzlibrary{arrows}
\usepackage[titletoc]{appendix}
\usepackage{breakcites}
\usepackage{authblk}
\usepackage{dsfont}
\usepackage{placeins}

\usepackage{scalerel}
\usepackage{natbib}
\usepackage{etoolbox}
\makeatletter
\patchcmd{\@makefntext}{\insertfootnotetext{#1}}{\insertfootnotetext{\scriptsize#1}}{}{}
\makeatother

\usepackage[normalem]{ulem} % to use \sout

\newcommand{\cut}[1]{{}}

\newcommand{\vp}{{\mathbf{p}}}

\newcommand{\vH}{{\mathbf{H}}}

\newcommand{\vR}{{\mathbf{R}}}

\newcommand{\EE}{\mathbb{E}}

\newcommand{\PP}{\mathbb{P}}

\DeclareBoldMathCommand\balpha{\alpha}
\DeclareBoldMathCommand\bdelta{\delta}
\DeclareBoldMathCommand\bsigma{\sigma}

\newcommand{\bc}{\begin{center}}
\newcommand{\ec}{\end{center}}

\newcommand{\bdm}{\begin{displaymath}}
\newcommand{\edm}{\end{displaymath}}

\newcommand{\beq}{\begin{equation}}
\newcommand{\eeq}{\end{equation}}

\newcommand{\bfl}{\begin{flushleft}}
\newcommand{\efl}{\end{flushleft}}

\newcommand{\bt}{\begin{tabbing}}
\newcommand{\et}{\end{tabbing}}

\newcommand{\beqn}{\begin{eqnarray}}
\newcommand{\eeqn}{\end{eqnarray}}

\newcommand{\beqs}{\begin{align*}} % no equation numbers
\newcommand{\eeqs}{\end{align*}}  % no equation numbers

\newcommand{\eqs}[1]{\begin{align} #1 \end{align}}
\newcommand{\eqss}[1]{\begin{align*} #1 \end{align*}}

\newcommand{\dd}{\mathrm{d}}
\newcommand{\pap}[2]{ \frac{\partial #1}{ \partial #2} }

\newtheorem{thm}{Theorem}[section]

\newtheorem{prop}[thm]{Proposition}

\newtheorem{coro}[thm]{Corollary}

 \newenvironment{pf}{{\bf Proof:}}{\hfill\rule{2mm}{2mm}}

\newcommand{\ba}{\left[ \begin{array}}
	\newcommand{\ea}{\\ \end{array} \right]}
\newcommand{\Var}{\mathrm{Var}}

%% file: sec0_abstract.tex
\begin{abstract}
\noindent
We study the implications of selling through a voice-based virtual assistant. The seller has a set of products available and the virtual assistant dynamically decides which product to offer in each sequential interaction and at what price. The virtual assistant may maximize the seller's profits; it may be altruistic, maximizing total surplus; or it may serve as a consumer agent maximizing the consumer surplus. The consumer is impatient and rational, seeking to maximize her expected utility given the information available to her. The virtual assistant selects products based on the consumer's request and other information available to it (e.g., consumer profile information) and presents them sequentially. Once a product is presented and priced, the consumer evaluates it and decides whether to make a purchase. The consumer's valuation of each product comprises a pre-evaluation value, which is common knowledge to the consumer and the virtual assistant, and a post-evaluation component which is private to the consumer. We solve for the equilibria and develop efficient algorithms for implementing the solution. In the special case where the private information is exponentially distributed, the profit-maximizing total surplus is distributed equally between the consumer and the seller, and the profit-maximizing ranking also maximizes the consumer surplus. We examine the effects of information asymmetry on the outcomes and study how incentive misalignment depends on the distribution of private valuations. We find that monotone rankings are optimal in the cases of a highly patient or impatient consumer and provide a good approximation for other levels of patience. The relationship between products' expected valuations and prices depends on the consumer's patience level and is monotone increasing (decreasing) when the consumer is highly impatient (patient). Also, the seller's share of total surplus decreases in the amount of private information.
We compare the virtual assistant to a traditional web-based interface, where multiple products are presented simultaneously on each page. We find that within a page, the higher-value products are priced lower than the lower-value product when the private valuations are exponentially distributed. This is because increasing one product's valuation increases the matching probability of the other products on the same page, which in turn increases their prices of other products. Finally, the web-based interface generally achieves higher profits for the seller than a virtual assistant due to the greater commitment power inherent in its presentation.

\end{abstract}

%% file: sec0_Intro.tex
\section{Introduction}
Electronic communication and commerce are in the early stages of a transition from the web and app eras, which were based on a screen- and keyboard-based user interface, to the use of natural language, primarily in the form of voice interfaces that are natural, fast and convenient for consumers.  With speech recognition reaching the point where natural language can be interpreted correctly more than $95\%$ of the time, most of today's operating systems come with voice interactivity across multiple devices ranging from phones and laptops to smart speakers and TVs. Amazon’s Echo family, Google Home and  Alibaba's Tmall Genie are examples of virtual assistant (VA) devices, and Amazon’s Alexa, Google’s Assistant, Microsoft’s Cortana and Apple's Siri are examples of voice-based user interfaces. The U.S. household penetration of VA devices was 32\% in 2019 and it is expected to increase to 51\% by 2022 (\cite{Forrester2019SmartHomeDevices}).
Further, these devices are expected to be integrated into cars, home electronics and consumer electronics devices that can be connected to platforms such as Amazon's Alexa. These new devices and interfaces offer a new way for consumers to access online information and increasingly, to engage in commerce. While the use of virtual assistants in electronic commerce is still in its infancy, it promises to become an important gateway to the marketplaces of the future, with more than sixty percent of marketers being optimistic about voice commerce compared to 11\% who are pessimistic (\cite{VoiceBot2019StateOfVA}).

\cite{Juniper2019Digital} estimated that there were more than 4 billion digital voice assistants worldwide in 2000, a number it expected to increase to 8.4 billion by 2024 due to the introduction of voice assistants in new devices including wearables, smart home, and TV devices. Voicebot.ai's Smart Speaker Consumer Adoption reports found that 15\% of U.S. smart speaker owners made purchases by voice on a monthly basis in 2018, up from 13.6\% in 2017 
(\cite{voicebot2019SSCCAR}, p. 17). Moreover, 20\% of consumers already shopped using smart speakers, and 44\% would make new or additional purchases if the interface was supported by more brands in 2019 (\cite{adobe2019}).

The voice interface offers speed and convenience but it also changes the nature of the interaction between the consumer and the marketplace. When using a virtual assistant, instead of searching and browsing a list of products within a page, the consumer uses natural language to specify what she is looking for and the virtual assistant presents products one by one through voice, based on the consumer’s requirements and other information. If the consumer rejects the first offer, the virtual assistant presents the next product, and so on. Thus, unlike traditional online channels where the consumer is largely in control of the search process, the sequential nature of voice response gives the virtual assistant a greater degree of control.

In this paper we study the implications of these new interfaces for product sales. We consider a seller with a set of products available for sale. The VA decides which product to offer in each sequential interaction and at what price to maximize the seller's expected profit, the total surplus or the consumer surplus. The consumer is rational, seeking to maximize her expected utility given the information available to her. The consumer has a limited attention span. If the consumer's search time exceeds her attention span, she leaves the market without making a purchase. When the consumer is active, the VA selects products and presents them to the consumer one by one. Once a product is presented to the consumer, she takes some time to evaluate it and decide whether to make a purchase. The consumer’s valuation of each product comprises two parts: a pre-evaluation component which is common knowledge to the consumer and the VA, and a post-evaluation component which is private to the consumer. Based on a comparison of value and price, the consumer decides whether to buy the current product or request another product, taking into account upcoming product opportunities as well as her limited purchase window.

% 1. equilibria and info asymmetry, different objective 
We first solve for the equilibria obtained when the consumer maximizes her expected utility and the VA  maximizes the seller's expected profit. We find efficient algorithms for solving the problem and compare them to the optimal results. We examine the effects of information asymmetries on the outcomes---pricing, ranking and surplus allocation. We consider the effects of incentive misalignment and find the equilibrium results for an altruistic VA (which maximizes the total surplus) and a consumer agent (which maximizes the consumer surplus). We study the effects of the distribution of private valuations and then compare the virtual assistant to a traditional web-based interface, where multiple products are presented simultaneously on each page. We find that the platform's surplus share is higher for the web-based interface. We also obtain closed-form results for the case where the private valuations are exponentially-distributed.

% under certain technical conditions the platform will rank products in a way which is socially-optimal, and we explore what happens under different distributions.

%% file: sec1_Literature.tex
\subsection{Literature review}

% Literature recognizes the potential 

% **** 1. Review of papers on VA， VA alone ****

Although the use of VAs in shopping is still an emerging phenomenon, the academic literature (e.g., \cite{kumar2016research}, \cite{rzepka2020another}) agrees with practitioners on its disruptive potential and expected growth. This literature is largely empirical or descriptive, focusing on the adoption, benefits and implications of VAs. \cite{kumar2016research} provide a taxonomy of intelligent agent technologies as well as a framework for studying their adoption. \cite{nasirian2017ai}  study the drivers of VA adoption by consumers, identifying interaction quality as a key driver of trust which (together with personal innovativeness) influences the intention to use a VA. \cite{sun2019effect} provide a comprehensive analysis of data from Alibaba's VA, TMall Genie, showing that the VA increases shopper engagement and purchase levels. They distinguish between the effect of VAs on purchase quantity, which is pronounced for high-income, younger, and more active consumers, and the effect on spending amount, which is larger for low-income, younger, and less active consumers. 

Several papers discuss both the benefits and limitations of VAs. \cite{rzepka2020another} examine the benefits and costs that users expect and obtain from voice commerce through semi-structured interviews with Amazon Alexa users. They identify convenience, efficiency and enjoyment as key perceived benefits and limited transparency, lack of trust and technical immaturity as perceived shortcomings. \cite{jones2018voice} uses a case study approach to illustrate these tradeoffs and their implications for consumers and marketers. She points out the challenges associated with consumers' loss of control and privacy. 
\cite{kraus2019voice} study factors that influence consumer satisfaction in voice commerce vs. e-commerce. They find that convenience is a key driver of satisfaction for both e-commerce and voice commerce, with a larger impact on the latter. They also bring a cognitive information processing perspective (\cite{davern2012cognition}) to the analysis, with the lower richness of voice commerce requiring a larger cognitive effort than for e-commerce. \cite{mari2019voice} and \cite{marievolution} use interviews with brand owners and AI experts, coupled with an analysis of the functional characteristics of VAs, to study the implications of the technology for consumers and brands. They identify the importance of ranking algorithms, finding that ``the virtual assistant may reduce consumers’ visibility of product alternatives and increase brand polarization. In this context, product ranking algorithms on virtual assistants assume an even more critical role than in other consumer applications'' (p.8), as the sequential nature of choice limits visibility, product alternatives and consumer choice. Similar limitations are also discussed in \cite{rzepka2020another} and \cite{jones2018voice}. One of our objectives in this paper is to study theoretically the implications of the ranking effect in light of the limitations of the technology. 

The limitations of buying through a VA are due to two factors. First, choices are made sequentially, which significantly limits consumers' ability to recall earlier options and make direct comparisons with them. Second, the voice interface is more limited than visual interfaces, which increases consumers' cognitive effort and again limits choice. Both drivers have an adverse effect on consumers as they increase cognitive load. \cite{basu2019choosing} and the papers they survey compare the effects of simultaneous and sequential presentation on consumer behavior. This literature shows that in general, sequential presentation results in inferior decisions compared to simultaneous presentation. Viewing options sequentially makes it difficult to properly compare the current choice to previous ones, whereas simultaneous presentation allows comprehensive comparisons. \cite{kraus2019voice} review the literature on the adverse effects of voice commerce due to the cognitive limitations of the auditory interface. \cite{munz2019not} show through six experiments that information presented by voice is more difficult to process than the same information presented visually. As a result of these cognitive difficulties, consumers tend not to recall the earlier options they have reviewed. Indeed, research in psychology and consumer behavior finds in multiple settings that consumers sampling products sequentially are often likely to choose the last product presented to them (e.g., \cite{de2003order}, \cite{bullard2017holding}). \cite{biswas2014making} argue that when consumers sample sensory-rich products, they may choose the first or the last product sampled depending on the degree of similarity across sensory cues. We expect that in the case of a VA, where the consumer decides when to stop searching, she is even more likely to select the last product considered when making a purchase.     

% Another important finding is that under sequential presentation, consumers tend not to recall the earlier options they have reviewed. \cite{bullard2017holding} find that under a sequential presentation, consumers are likely to choose the last item that was presented to them. \cite{de2003order} and \cite{bullard2017holding} suggest that the consumer tend to make decision without recalling previously presented items when items are sequentially presented. This effect was greatly reduced or was completely eliminated when the items are presented simultaneously. It is likely to be even stronger with a voice interface, which increases the consumer's cognitive effort.  **** I HALF REWROTE THE PARAGRAPH BUT IT NEEDS MORE WORK -- there's too much repetition and too little information here ****  
 
%END EMPIRICAL RESULTS TO BE MOVED UP *****
% ********
%**** BRING HERE ALSO THE CONCLUSIONS ON NO RECALL, ETC. 

The existing VA literature is largely empirical. As researchers (e.g., \cite{mari2019voice}, \cite{marievolution}) have noted, ``providing structure and guidance to researchers and marketers in order to further explore this emerging stream of research (VA) is fundamental" (p.8). In this paper, we aim to narrow the gap between theory and practice by modeling the VA buying process, determining how a VA should price and rank products to maximize alternative objective functions given rational consumer choice, studying the implications of the resulting equilibria, and comparing the outcomes obtained through a VA to those obtained from a web interface. 

%2. Analytical: assortment without ranking effect 
Analytically, our model shares some aspects (in particular, consumer impatience) with \cite{izhutov2018pricing}, who consider a two-sided marketplace for services such as tutoring and derive pricing algorithms to maximize social surplus or seller profits. In \cite{izhutov2018pricing}, the seller can only control prices (there is no ranking problem) and consumer behavior is exogenous. In contrast, our model studies a VA selling physical products and controlling both the prices and the order of presentation to the consumer. Our optimization problem is also analytically related to the retail assortment planning problem in the Operations Management literature. This literature may be classified into three overlapping groups: \textit{(i)} traditional assortment planning analysis with no ranking effect, which is more foundational in nature; \textit{(ii)} ranking through an auction or similar means, where an intermediary platform designs a mechanism that ranks offers based on suppliers' bids or similar information signals; and \textit{(iii)} more complex (mostly two-stage) models of assortment ranking.

The traditional assortment planning literature (group \textit{(i)}) derives algorithms to compute the optimal assortment for a retail operation. The problem may be addressed in either static or dynamic setting. \cite{kok2008assortment} provide an extensive review of the static assortment planning literature. \cite{mahajan2001stocking} and \cite{honhon2010assortment}
 study the optimal assortment when consumers can only choose among the products that are still in stock.  \cite{golrezaei2014real} and \cite{bernstein2015dynamic}
 consider the problem of dynamically customizing assortment offerings based on the preferences of each consumer and the remaining product inventories. % In both papers, the dynamics are driven by the different consumer types and the remaining inventory levels.
 Motivated by fast fashion operations, \cite{caro2014assortment} study how to release products from a fixed set into stores over multiple periods, taking into account the decay in product attractiveness once presented in the store. \cite{davis2015assortment} study the assortment planning problem for a seller that sequentially adds products to its assortment over time, thereby monotonically increasing consumers’ consideration sets. In this setting, the profit margin is exogenously given and the consumer is myopic. \cite{davis2015assortment} derive an approximation algorithm that achieves at least $(1-1/e)$ of the optimal revenue. \cite{saban2019procurement} consider the mechanism design problem for a procurement agency that selects suppliers or product assortments made available to consumers. Suppliers may compete \textit{for} the market (entry/assortment competition) and \textit{in} the market (price competition within the assortment). They characterize the optimal buying mechanisms, showing how restricting the entry of close substitutes may increase price competition and consumer surplus. 
 
 \cite{ferreira2019assortment} introduce information effects to the assortment planning literature, considering the impact of concealing products which are in the full product catalog from consumers in an attempt to induce them to buy additional products later. They compare the case where the seller presents the entire assortment for the entire selling season (similar to a traditional department store or a web-based catalog) to the case where the seller intentionally introduces products one at a time (fast-fashion being a canonical example). The difference between the seller's expected profits in the latter vs. the former case is the value of concealment. When consumers are non-anticipating (i.e., make decisions a period at a time ignoring the future), the retailer successfully induces them to buy more products, which increases its profit. When consumers are strategic and anticipate the retailer's future actions, the value of concealment is ambiguous. 
 While most of the paper focuses on the value of concealment for a given ranking and assortment, (\cite{ferreira2019assortment}, Corollary 1 and 2) also show that under a special set of circumstances, the seller will induce the consumer to buy more by concealing the more valuable products so she will buy them later. This happens when the consumer is non-anticipating and either \textit{(i)} the prices (assumed exogenous throughout the paper) are the same and the valuation distributions follow a stochastic dominance relationship, or \textit{(ii)} the valuations have the same distribution which has an increasing failure rate. \cite{ferreira2019assortment} thus provide a nice model of information disclosure as a tool sellers can manipulate to their advantage. 
 
% 3. Analytical: Ranking effect
As discussed above, the product ranking decision is particularly important when selling through a VA (cf. \cite{mari2019voice}, \cite{marievolution}). This problem may be related to the vast literature studying how a search engine should rank paid ads. These models are typically three-sided, involving consumers, advertisers and the search engine, and they focus on a tradeoff between relevance and revenue (see, e.g, \cite{agrawal2020optimization} for a review). \cite{l2017revenue} bring this tradeoff to the assortment planning domain considering multiple sellers selling one item each through a platform. \cite{l2017revenue} summarize the behavior of consumers using a click-through rate function that represents the purchase probability of each item as a function of its position and intrinsic characteristics. The platform maximizes expected revenue by calculating for each item a score that balances its revenue and relevance and ranking them by decreasing scores. \cite{chu2020position} study a three-sided market involving a consumer and multiple suppliers with a platform in-between. The platform's objective is to maximize a weighted average of the suppliers' surplus, consumer surplus, and platform revenue. They study a position auction (in the spirit of search engine position auctions) to sell the first $k$ available slots to $n$ suppliers. They rank suppliers based on a surplus-ordered ranking (SOR) mechanism that considers suppliers' contributions to the objective function (realized for the first $k$ and expected for the remaining suppliers), and show numerically that this mechanism is near-optimal. From the consumers' point of view, products are ex-ante homogeneous, which allows \cite{chu2020position} to show that, in a simpler setup with no auction, the platform's optimal ranking is based on realized SOR with full information and expected SOR with no information. 

Ranking decisions may be made by the consumer, by the seller, or by some combination of the two. The classic framework for ranking by the consumer is due to \cite{weitzman1979optimal}, who proposes an elegant solution to a consumer choice model he calls ``Pandora's Box'': a consumer faces $N$ closed boxes, where box $i$ has a prize whose (unknown) value comes from a known distribution, and the cost to open box $i$ is $c_i$. The consumer has to decide in what order to open and inspect the boxes (to learn the realized value of the prize inside), and when to stop opening and take home the best prize, allowing perfect recall. The solution ranks boxes by a reservation price index, with the consumer stopping when the maximum sampled reward exceeds the reservation price of every remaining closed box. A few papers in the assortment planning literature use the idea of a two-stage consumer choice process, where the consumer first forms a consideration set in the spirit of \cite{weitzman1979optimal}, and in the second stage selects the product within that consideration set. This is an effective model for a web-based seller but not for a VA, where the order of inspections is determined by the VA, not by the consumer. \cite{wang2018impact} study assortment planning and pricing where the consumer follows a two-stage model with homogeneous search costs. The consumer includes products in her consideration set by comparing their incremental net utility to the search cost. In the second stage, the consumer uses the Multinomial Logistic (MNL) model (assuming Gumbel random valuations) to select from the consideration set. \cite{wang2018impact} develop an algorithm to calculate the optimal assortment and prices. They find that without a search cost, all products have the same price---a consequence of the MNL assumption.
\cite{negin2017} develop a two-stage model with heterogeneous search costs which are increasing in the item's position. Prices are exogenous and consumers follow a process similar in spirit to \cite{weitzman1979optimal} to form a consideration set. The final choice is then made following an MNL choice model similar to \cite{wang2018impact}. They show that the problem is NP hard and propose a polynomial-time solution algorithm. They find that even though ranking products in descending order of intrinsic utilities is suboptimal, it achieves a multiplicative approximation factor of 1/2 and an additive factor of 0.17. 

The web-interface model developed by \cite{gallego2018approximation} also adopts the two-stage consumer choice approach: the consumer first forms a consideration set (all products in the first $k$ pages) according to her type (which determines $k$) and then follows the MNL model to select a product within the consideration set. The seller knows the distribution of consumer types and maximizes its expected revenue over rankings and prices. \cite{gallego2018approximation} recommend a descending ranking in value gaps (net utilities when  products are sold at their unit wholesale costs). They show that the price markups are the same for all products on the same page, and are increasing in the page index. \cite{aouad2015assortment} propose a two-stage model outside the \cite{weitzman1979optimal} setup. Instead, different consumer types have exogenous consideration sets and product preferences, and the platform determines the overall assortment from which consumers select their preferred product. \cite{aouad2015assortment} formulate the problem as one of maximizing expected revenue over a graph and derive a recursive algorithm for solving it.

% \subsubsection{Comparison}
To our knowledge, our model is the first that combines optimal ranking and pricing decisions by a seller under strategic consumer choice in the setting of a VA. Because of our different objectives, both our model structure and our results are significantly different from those in the assortment optimization literature. Relative to the traditional assortment planning literature, these differences are obvious, as our model is designed to jointly determine the optimal ranking and pricing for the particular setting of a VA. This is why ranking and presentation are at the core of our model as opposed to most traditional assortment planning models (group \textit{(i)} in our classification of the literature). In contrast to the myopic consumer assumed in most traditional models, we model a rational, forward-looking consumer. As a result, the temporal dynamics are the equilibrium outcome of the interactions between the seller and the consumer, unlike traditional (group \textit{(i)}) assortment planning models where the dynamics are driven by the seller's inventory level. These and other differences are to be expected as this literature was not designed to address the issues we focus on in this paper.  

Our model is closer in structure to those of ~\cite{l2017revenue},  \cite{chu2020position}, \cite{wang2018impact}, \cite{negin2017}, \cite{gallego2018approximation}, \cite{aouad2015assortment} and  \cite{ferreira2019assortment}, which we reviewed above. The key differences between these models and ours are summarized below.
\begin{enumerate}

    \item \textbf{Pricing.} Most (but not all) of the above papers focus on assortment planning per se and therefore take product prices as given (\cite{ferreira2019assortment}, \cite{l2017revenue}, \cite{chu2020position} \cite{negin2017}, \cite{aouad2015assortment}), or use the MNL assumption (\cite{wang2018impact} \cite{gallego2018approximation}). As is well-known, the MNL assumption greatly simplifies the price optimization problem, however it is restrictive and is designed to lead to the same price or margin for all products (\cite{anderson1992discrete}). Indeed, \cite{gallego2018approximation} find the same price margin within a same page and \cite{wang2018impact} find the same price for all except possibly one product. In our setting, the seller has two levers at hand, pricing and ranking, and our model allows it to take full advantage of both, which enables us to derive key insights on their interaction and impact. 

    \item \textbf{Consumer choice and ranking model.} The literature covers multiple consumer choice and ranking models, all of which significantly differ from ours. \cite{wang2018impact} and \cite{aouad2015assortment} differ from \cite{negin2017} in that the former focus on two-stage assortment optimization without an actual position (i.e., ranking) effect whereas the latter considers position-dependent consumer search costs. In all three papers, the consumer has the freedom to choose which product to review---an appropriate structure for a store or a web-based seller, but not for a VA, where the VA, rather than the consumer, makes this choice. In \cite{gallego2018approximation}'s two-stage model, the consumer views \emph{all} the products in the first few pages without deciding when to stop, as the consumer type determines the number of pages (and products) she views. Similar to \cite{negin2017} and \cite{wang2018impact} and unlike our setting, the final choice is made by the consumer based on the MNL choice model. \cite{l2017revenue} and \cite{chu2020position} model a platform in the spirit of the search engine literature, with the consumer choice problem not being influenced by the platform's strategy, or being modeled indirectly. \cite{chu2020position}'s three-sided model involves a platform, suppliers and consumers. The consumer's decisions need not depend on the platform’s ranking strategy as all products are ex-ante homogeneous from the consumer's point of view. \cite{l2017revenue} model the consumer choice problem indirectly through a click-through-rate function that combines the effects of position and product characteristics. Unlike the foregoing papers, \cite{ferreira2019assortment} consider information strategies designed to increase the number of units bought by a consumer over a selling season. For the most part, they consider the implications of a \emph{given} ranking and assortment on the value of concealment. Two of their Corollaries show show that under a special set of circumstances, a particular ranking (ascending order in product valuations) is optimal. The consumer choice model leading to this result is entirely different from ours (an infinitely patient non-anticipating consumer buying multiple units) as are the assumptions (same prices or same valuation distributions) and results, differences which are driven by their different objective (to study the value of concealment).

%In \cite{chu2020position} and \cite{ferreira2019assortment}, **** GROUP THEM AND FOCUS ON DIFFERENCES. 
    
   % In the papers mentioned above, ranking is either \textit{(i)} search-engine-like where suppliers bid for product position, as in \cite{saban2019procurement} and  \cite{chu2020position}), or \textit{(ii)} based on a Weitzman-like consideration set determined by consumers (\cite{negin2017} \cite{wang2018impact}  or \textit{(iii)} by consumer type  (\cite{gallego2018approximation}, \cite{aouad2015assortment} ), \textit{(iv)} by a given click function (\cite{l2017revenue}), or \textit{(v)} by platform to maximize its objective function (\cite{ferreira2019assortment}). These are appropriate structures for a web-based platform, but not for a VA, where the consumer does not have the freedom to choose which item to review. Instead, they must follow the sequence determined by the VA.  
    
    \item \textbf{Recall.} All the above papers assume that consumers have perfect recall. Whereas this assumption is reasonable for a traditional web-interface (at the page level), as discussed above, choice under a sequential presentation (especially with a voice interface, which further increases cognitive load) is better characterized by no recall. It will be interesting to examine the effects of imperfect recall, which covers both assumptions as special cases, in future work.

    \item \textbf{Comparison of VA to web interface.} An important issue studied in our paper is the comparison between selling through a web interface and selling through a VA. Naturally, the papers cited above were not designed to provide such a comparison.
    %\cite{gallego2018approximation}'s model, reviewed above, provides a good approximation for the page-by-page nature of a web interface. In their setting, the consideration set is formed exogenously (by consumer type) and is not affected by the seller's pricing and ranking decisions. These assumptions are suitable for modeling a web-interface, however, due to the higher cognitive load in VA setting, their model may not be able to illustrate the direct comparison between two interfaces. Contrast to their setting, the consumer's consideration set/stopping time/threshold policy etc. As we are going to see later, those modeling difference drives very different results/insight.
\end{enumerate}

% Our sequential presentation structure is closer to \cite{ferreira2019assortment}, which has a different objective that drives different modeling assumptions and hence results. **** THIS IS TOO SUPERFICIAL ****We show, under a similar setting (sequential presentation, consumer with infinite patience but no recall), that the optimal ranking is to hold the most valuable product (in terms of value minus price) to later periods, even when the consumer is forward-looking. 

The differences between the objectives and modeling choices of our paper vis-a-vis the foregoing papers also drive key differences in the results: 

\begin{enumerate}
    \item \textbf{Optimal pricing}. Most of the above papers take prices as given, \cite{wang2018impact} and \cite{gallego2018approximation} being the exceptions. The MNL consumer choice model used in both papers naturally gives rise to pricing results that are significantly different from ours. \cite{wang2018impact} find that either all products are priced the same, similar to the MNL literature, or all but one product are priced the same (the lowest value product may be priced differently so it will enter the consideration set without affecting the prices of higher-valued products). 
   \cite{gallego2018approximation} find that with a web interface, all products within each page have the same price margin, a result which is mainly driven by the MNL assumption. In contrast, we find---for our web interface model---the counter-intuitive result that within a page, prices are typically decreasing in product valuations (or margins), which means that higher-valued products tend to have a lower price. Higher-valued products are priced lower to increase the matching probability of the entire page, thus allowing lower-valued products to set a higher price. This insight does not apply to  \cite{gallego2018approximation}'s model, where a given consumer can simultaneously inspect all products on the pages she has access to. \cite{gallego2018approximation} also find that across web pages, the price margin is increasing in the page index. In contrast, we find that the reverse may also happen depending on problem parameters such as the amount of private information and the consumer patience level. 
     
    \item \textbf{Optimal ranking}.
    Among the papers where the \emph{seller} solves a ranking problem, one suggests an approximation algorithm with performance guarantees (\cite{negin2017}), and two characterize the optimal results (\cite{chu2020position}, \cite{l2017revenue}). \cite{negin2017} develop an approximate ranking algorithm and show that the intuitive ranking (descending in product utility), though not optimal, has a performance guarantee. In \cite{chu2020position}, the optimal ranking is descending in the product's contribution to the platform’s (weighted) objective function. In \cite{l2017revenue}, the optimal ranking is decreasing in product scores that summarize each product's contribution to expected revenue. In all three papers, products are ranked by a decreasing score that quantifies their contribution to the objective function. Due to the flexibility of our model, we find the optimal ranking may be an arbitrary permutation of the available products depending on the consumer patience level and other problem parameters. Further, under certain conditions the optimal ranking is actually \emph{increasing} in product valuations.  \cite{wang2018impact} and \cite{aouad2015assortment} consider the case where the seller provides an assortment to the consumer, and the consumer ranks the products to form a consideration set and then selects the best item within the consideration set. The optimal ranking in their model is descending in the consumer utility (\cite{wang2018impact}) or in the consumer's preference list (\cite{aouad2018approximation}). In our paper, the optimal ranking for a profit-maximizing seller is highly-dependent on the consumer's patience level. In particular, the optimal ranking is descending in valuations for a highly-impatient consumer and ascending in valuations for a sufficiently patient consumer. This is intertwined with our model's pricing decisions: an ascending order, for example, allows the seller to induce the consumer to stay and pay a higher price for lower-ranked products. This strategy, however, can work only when the consumer is sufficiently patient. Even for a consumer agent that maximizes consumer surplus, the optimal ranking is still ascending in valuations when the consumer is patient as the consumer benefits from viewing more products. 
    
    \item \textbf{Comparison of alternative objective functions}. Most of the above papers (\cite{ferreira2019assortment}, \cite{l2017revenue}, \cite{gallego2018approximation}, \cite{aouad2015assortment}, \cite{ferreira2019assortment}) consider a profit- (or revenue-)maximizing seller or platform, \cite{chu2020position} and \cite{negin2017} being the exceptions. \cite{chu2020position}'s objective function is a weighted average of the platform revenue, supplier surplus and consumer surplus. They find that the optimal ranking for a profit-maximizing platform is descending in product prices; for maximizing supplier surplus, it is descending in the seller valuations; and for maximizing consumer surplus, it is descending in the consumer's utility. \cite{negin2017} also develop an algorithm for maximizing consumer welfare. Both papers provide algorithms to approximate the optimal ranking and, unlike our paper, do not focus on structural differences in the results obtained under alternative objective functions. In our paper, we study the differences in surplus allocations and strategies for VAs that maximize seller profits, overall surplus (an altruistic VA) or consumer surplus (a consumer agent). Our specification allows us to study how the surplus allocation depends on the distribution of the private signals and on the consumer's patience level. Interestingly, when the private valuation distribution is exponential, a profit-maximizing seller always extracts half the total surplus. We also find that the seller makes zero profit when the VA is altruistic or acts as a consumer agent. Again, unlike \cite{chu2020position} and \cite{negin2017}, we can also study the effect on product prices. We find that an altruistic seller (maximizing total surplus) follows exactly the same ranking and pricing strategy as a consumer agent. The strategies of a a seller-profit-maximizing VA reflect its market power and its ability to manipulate the presentation to its advantage. Nevertheless, for both highly patient and highly impatient consumers and exponentially-distributed private valuations, the optimal rankings under all three objective functions are the same.

     % Due to the consumer's search behavior, the consumer will include products one by one, in a descending order w.r.t its value minus cost, until the marginal benefit of including one extra product exceeds the search cost. In order to make consumer include the lower valued product, the seller may either lower the prices for all higher valued products (same-price policy), or only lower the price for this particular product (quasi-same-price policy).
\end{enumerate}

%% file: sec1_Overview.tex
\subsection{Paper Overview}
The plan of the paper is as follows. Following this Introduction, Section \ref{sec:model_formulation} presents our model. The pricing problem for a given ranking is solved in Section \ref{sec:equlibrium_pricing}. Section \ref{sec:optimal_ranking} addresses the optimal ranking. Section \ref{sec:implication} discusses the implications of using a VA, Section \ref{sec:web_vs_va} compares the VA to a traditional web-interface and Section\ref{sec:conclusion} offers our concluding remarks.
\\

%% file: sec2_ModelExp.tex
\section{Model}\label{sec:model_formulation}
% model description/ setting
We consider a virtual assistant serving a seller that sequentially offers products to consumers and prices them dynamically. The seller (through the VA) selects from $N$ products $i=1,2,\cdots, N$ that are available for sale, presenting and pricing them one at a time. When the seller sells product $i$, it incurs a cost $c_i$ which includes payments to the vendor, logistics and shipping (if applicable) costs, etc., and it pockets the difference between the price it charges, $p_i$, and its cost $c_i$. An arriving consumer specifies what she is looking for and the VA then presents the available products sequentially at its discretion. When a product $i$ is presented, the seller prices it dynamically at $p_i$ based on the information available to it. The consumer evaluates the product and decides whether to buy it or move on to the next product (in which case product $i$ becomes unavailable). Each product presentation and evaluation lasts an exponential amount of time with rate $\tau_p$ and expected value $1/\tau_p$. The consumer's overall search lasts for an exponential amount of time with rate $\tau_c$ and expected value $1/\tau_c$. The consumer leaves the system following this search time or after she  exhausts all $N$ available products without making a purchase. All of the evaluation and search times are independent. We denote by $\rho= \frac{\tau_p}{\tau_p+\tau_c}$ the consumer's patience parameter---the probability that for a given product, the consumer's evaluation is completed before she leaves the system (so she can buy it if she chooses to). The seller uses the VA to rank and dynamically price each product so as to maximize its expected profit. With this objective function, the VA is subservient to the seller. We also consider other objectives (maximizing total surplus and maximizing consumer surplus) in Section \ref{subsec:diff_obj}. As consumers arrive sequentially, we can perform the analysis for one consumer at a time.

When a consumer arrives, she provides information about the product she is looking for (e.g., in the form of a search query). This information, along with consumer profile information, is converted by the seller to an estimated base valuation $v_i$ for each product $i$. Once the consumer has evaluated a product presented to her, she values the product at $v_i+\epsilon_i$, where $\epsilon_i$ is the private information unobservable by the seller. When product $i$ is priced at $p_i$, the consumer’s net utility from purchasing it is thus $ u_i = v_i -p_i + \epsilon_i$. We assume the $\epsilon_i$ are i.i.d. random variables sampled from a common distribution $F(\cdot)$ with positive support.  

The consumer aims to maximize her expected net utility. We show later that in equilibrium, the consumer will buy product $i$ if and only if its net utility $u_i$ is above a threshold level that balances the net utility of product $i$ against the opportunity to buy a product yielding a higher net utility in the future. We call this form of consumer policy ``threshold policy''. For simplicity, we assume no discounting (as is well known, a discount factor may be incorporated in $\tau_c$). If product $i$ were sold at cost, its
% \commwb{justification of why consider $p_i = c_i$? Say in a perfect competition setting}, its expected \emph{ex ante} 
value to the consumer would be $v_i + \EE(\epsilon) - c_i.$ We call $v_i +\EE(\epsilon) - c_i $ the \emph{value margin} of product $i (i=1,2, \ldots, N)$ and assume for simplicity that the $(v_i - c_i )$ are non-positive and $\epsilon$ is positively-supported.
% **** WHAT WOULD BE THE IMPLICATIONS OF CALLING $v_i + \EE(\epsilon) - c_i$ value margin and working with that? ****%Relaxing this assumption may result in corner solution (see Appendix \ref{app:assumption_negative}). **** Add Later ****

In summary, the seller decides on a product ranking and dynamic pricing policy to maximize its objective function knowing all $v_i, c_i~(i = 1,2,\cdots, N)$ and which products have been rejected by the consumer. The consumer dynamically decides on whether to accept or reject each offer, and the process continues until the consumer departs. Our notation is summarized in appendix \ref{app_notation}.

% Decide later. 07/15

\subsection{Strategies and equilibrium}

Each party (the seller and the consumer) hold beliefs about the other’s strategy, and each optimizes its own objective function taking the other’s strategy as given. In equilibrium, these beliefs are consistent.
% \begin{absolutelynopagebreak}
% \begin{definition}{\textbf{Strategies and equilibrium}}\\
The seller's strategy is defined by two $n$-dimensional vectors:
    \begin{enumerate}[label=(\roman*)]
        \item  a permutation $\bsigma = (\sigma_1, \cdots, \sigma_N)$  specifying the ranking of products for presentation by the seller, and
        \item a price vector $\vp = (p_{\sigma_1}, \cdots  ,p_{\sigma_N})$  specifying the price assigned by the seller to each product.
    \end{enumerate}
 The consumer's strategy is defined as a mapping from the history up to time $t$, $\vH_t$, to $\{\text{accept, reject\}}$.

In equilibrium,
    \begin{enumerate}[label=(\roman*)]
        \item given the price $\vp$ and ranking $\bsigma$, the consumer’s acceptance strategy at each time $t$ is optimal; and
        \item given the consumer’s strategy, the rank-price combination $(\bsigma, \vp)$ maximizes the seller's objective function.
        \end{enumerate}

% \end{absolutelynopagebreak}
\noindent Next, we show that in equilibrium, the consumer's optimal strategy must have the threshold policy structure. Assume, without loss of generality, that for a given seller strategy $(\bsigma, \vp)$, $\bsigma = (1,2,\cdots, N)$ by renumbering the products. Let $V_i^c(\epsilon_i)$  be the expected consumer utility from round $i$ until the consumer departs (possibly purchasing a product) when she observes a private value of $\epsilon_i$. Let $V_i^c= \EE V_i^c(\epsilon_i)$ where the expectation is taken over $\epsilon_i$. Then, if the consumer decides to buy in round $i$, she'll get $ v_i – p_i+ \epsilon_i $. Otherwise, the consumer waits for the next round and receives an expected utility of $\rho V_{i+1}^c$.  It follows that
\[ V_i^c (\epsilon_i) = \max \{v_i – p_i+ \epsilon_i, \rho V_{i+1}^c\}.\]
The consumer will purchase in round $i$ if and only if $u_i=v_i – p_i+ \epsilon_i \geq \rho V_{i+1}^c$. Thus, the consumer will choose a threshold policy: $u_i \geq \delta_i: =\rho V_{i+1}^c$, where $\delta_i$ is the consumer's optimal threshold in round $i$. We have:
\begin{proposition}{\textbf{\emph{(Consumer threshold policy).}}} \label{prop1}
For any given $(\bsigma,\vp)$, the consumer’s best response is a threshold policy $\bdelta$.
\end{proposition}
\smallskip
An equilibrium is thus defined by three $n$-dimensional vectors $(\bdelta,\bm{\sigma},\bm{p})$ such that: 
% *** I DON'T THINK THAT $V^p$ HAS BEEN DEFINED YET. ***
\begin{align}\label{eq:def_eqm_pp}
{\bdelta}&\ \textup{maximizes} \ \EE[V^c(\bdelta,\bsigma,\vp)]
\textup{ , and} \\ 
(\bsigma,\vp) &\  \textup{maximizes} \ \EE[V^p(\bm{\delta},\bsigma,\bm{p})],
\end{align}
where $V^p$ ($V^c$) is the expected seller (consumer) surplus.

%% file: sec3_EquilirbiumPrice.tex
\section{Equilibrium pricing} \label{sec:equlibrium_pricing}

We first solve the profit-maximizing pricing problem for a given ranking $\bsigma$, where the i.i.d. private valuations $\epsilon_i$ follow a general distribution $F$. We then consider the special case where they are exponentially distributed. In section \ref{subsec:diff_obj}, we examine what happens under alternative objective functions.

\subsection{Pricing for profit maximization} \label{subsec:va_pp_pricing}
%\\
For a given ranking $\bsigma$ (assuming, without loss of generality, $\bsigma = (1, 2, \cdots, N)$), the equilibrium prices and the consumer's optimal thresholds can be calculated by backward induction, yielding the following proposition. 

\begin{prop}{\emph{\textbf{(Profit-maximization for a given ranking with $\epsilon \sim F(\cdot)$).}}}\label{prop:pp_eqm_general}
Given $\bsigma= (1, 2, \cdots, N)$, the equilibrium $(\bdelta, \vp)$  satisfies the following recursion:
\begin{enumerate}[label=(\alph*)]
    \item In stage $N$,
    \begin{itemize}  
        \item[] The consumer's threshold is $\delta_N=0$. 
        \item[] The equilibrium price at stage $N$ is given by $$p^*_N = \arg\max_{p} \PP_N(p-c_N),$$ 
        and the equilibrium probability of purchase is given by $$\PP_N=1-F(p^*_N-v_N).$$
        \item[] Under the optimal price, the seller's expected profit in period $N$ is given by
        $$V^p_N=\PP_N(p^*_N-c_N),$$
        and the consumer's expected surplus in period $N$ is given by $$V^c_N = \PP_{N} \EE(\epsilon_N + v_N-p^*_N|\epsilon_N +v_N-p^*_N\geq\delta_N).$$
    \end{itemize}
    \item In stage $i=N-1, \cdots, 1,$
    \begin{itemize}
        \item[] The consumer's threshold is $\delta_{i}  = \rho V^c_{i+1}$.
        \item[] The equilibrium price in stage $i$ is given by
        $$p^*_i=\arg\max_{p} \PP_{i} (p-c_{i}) +  (1-\PP_{i})\rho  V_{i+1}^p, $$
        
        and the probability of purchase is given by $$\PP_{i}=1-F(p^*_i-v_{i}+\rho V_{i+1}^c).$$
        
        \item[] Under the optimal price, the seller's expected profit in period $i$ is given by
        $$V^p_i=\PP_{i} (p^*_i-c_{i}) +  (1-\PP_{i})\rho  V_{i+1}^p,$$
        and the consumer's expected surplus in period $i$ is given by
        $$V^c_{i}  =  \PP_{i} \EE(\epsilon_{i} + v_{i}-p^*_{i}|\epsilon_{i} + v_{i}-p^*_{i}\geq \delta_i ) + (1-\PP_{i})  \rho V_{i+1}^c.$$
    \end{itemize}

\end{enumerate}
\end{prop}
\noindent
In the special case where the private valuations are exponentially distributed, Proposition \ref{prop:pp_eqm_general} simplifies to:

\begin{coro}{\textbf{\emph{(Profit-maximization: Equilibrium for a given ranking with  $\epsilon \sim \exp(\alpha)$).}}}\label{cor:prop_pp_eqm_exp} Given $\bsigma=(1, 2, \cdots, N)$, the equilibrium $(\bdelta, \vp)$  satisfies the following recursion:
\begin{enumerate}[label=(\alph*)]
    \item In stage $N$,
    \begin{itemize}
        \item[] The consumer's threshold is $\delta_N=0$, the equilibrium price is  $$p^*_N=c_N+\frac{1}{\alpha},$$
        and the probability of purchase is given by
        $$\PP_N=\exp(\alpha(v_N-c_N)-1).$$
        \item[] Under the optimal price, the seller's expected profit and the consumer's expected surplus are both given by 
        $$V^p_N=V^c_N =\dfrac{1}{\alpha}\exp(\alpha(v_N-c_N)-1)=\dfrac{1}{\alpha}\PP_N$$
    \end{itemize}
    \item At stage $i=N-1,\dots,1$,
    \begin{itemize}
        \item[] The consumer's threshold is $\delta_i=\rho V^c_{i+1}$, 
        \item[] The optimal price is $$p^*_i=c_i+\frac{1}{\alpha}+\rho V^p_{i+1},$$
        and the probability of purchase is         $$\PP_i=\exp(\alpha(v_i-c_i)-1) \cdot \exp(-\alpha\rho(2V^p_{i+1}))$$ 
        % \commwb{since $V^c_{i+1}=V^p_{i+1} $, shall we simplify it by $2V^c_{i+1}$}  *** Yes, but for consistency, we should use p, not c *****
        \item[] Under the optimal price, the seller's expected profit and the consumer's expected surplus at stage $i$ is  
        $$V^p_i=V^c_i = \dfrac{1}{\alpha}\sum^{N}_{k=i}\rho^{k-i}\PP_k.$$
    \end{itemize}
\end{enumerate}
\end{coro}

\noindent By Proposition \ref{prop:pp_eqm_general}, the optimal price in the last stage is the ``monopoly'' price that maximizes the seller's expected profit in the single-product case. This is because in the last stage, the seller is effectively making a final "take-it-or-leave-it" offer to the consumer with no continuation option. In the exponential case ({Corollary \ref{cor:prop_pp_eqm_exp}}), the monopoly price in the last stage is independent of the last product's valuation, and  the price of the product offered in each stage $i$ is independent of its own valuation $v_i$ -- it depends only on the valuations of products $ i+1, \ldots N$ through the continuation value $\rho V^{p}_{i+1}$. The optimal price has a special structure -- it equals the monopoly price plus the future continuation value. This implies that products viewed in earlier stages receive a higher price markup to induce the consumer to view more products which, in turn, increases the expected private valuation of the product purchased and with it, the seller's expected profit.

Another interesting consequence of Corollary \ref{cor:prop_pp_eqm_exp} is that in the exponential case, for any given ranking, the seller's expected profit is equal to the expected consumer surplus. This implies that the ranking that maximizes the seller's profits also maximizes the consumer surplus: 
\begin{coro}
%{\textbf{\emph{(Profit-maximizing seller with exponential private valuations: Incentive alignment).}}}\\
When $\epsilon$ is exponentially distributed, the profit-maximizing pricing solution equates the seller's expected profit to the consumer surplus for any given ranking. Thus, the profit-maximizing ranking also maximizes the consumer surplus.
\end{coro}

In the next subsection, we solve the optimal pricing problem for a VA with different objective functions: \emph{(i)} a consumer agent that maximizes the expected consumer surplus, and \emph{(ii)} an altruistic VA that maximizes the expected total surplus. 

\subsection{Alternative objective functions} \label{subsec:diff_obj}

The results of Section \ref{subsec:va_pp_pricing} can be directly generalized to VAs with different objective functions. The VA may be altruistic, maximizing the total surplus $V^s =V^c + V^p$, or it may serve as a consumer agent, maximizing the consumer surplus $V^c$. As we show below, in both cases the seller extracts no surplus and each product is priced at cost:
\begin{prop}{\textbf{\emph{(Altruistic/Consumer agent VA: Equilibrium for a given ranking with $\epsilon \sim F(\cdot)$}):}}\label{prop:apca_eqm} Given $\bsigma= (1, 2, \cdots, N)$, the equilibrium $(\bdelta, \vp)$ satisfy the following recursion:
\begin{enumerate}[label=(\alph*)]    
\item In stage $N$,
    \begin{itemize}
        \item[] The consumer's threshold is $\delta_N=0$. 
        \item[] The equilibrium optimal price is $$p^*_N = c_N$$ 
        and the equilibrium probability of purchase is $$\PP_N=1-F(c_N-v_N).$$
        \item[] Under the optimal price, the seller's expected profit is 
        $$V^p_N=0,$$
        and the consumer's expected surplus is $$V^c_N = \PP_{N} \EE(\epsilon_N + v_N-c_N|\epsilon_N +v_N-c_N\geq\delta_N).$$
    \end{itemize}
    \item In stage $i=N-1, \cdots, 1,$
    \begin{itemize}
        \item[] The consumer's threshold is $\delta_{i}  = \rho V^c_{i+1}$.
        \item[] The equilibrium optimal price is 
        $$p^*_i=c_i$$
        and the equilibrium probability of purchase is $$\PP_{i}=1-F(c_i-v_{i}+\rho V_{i+1}^c).$$
        
        \item[] Under the optimal price, the seller's expected profit is 
        $$V^p_i=0,$$
        and the consumer's expected surplus is 
        $$V^c_{i} =  \PP_{i} \EE(\epsilon_{i} + v_{i}-c_{i}|\epsilon_{i} + v_{i}-c_{i}\geq \delta_i ) + (1-\PP_{i})  \rho V_{i+1}^c.$$
    \end{itemize}

\end{enumerate}

\end{prop}
By Proposition \ref{prop:apca_eqm}, both the altruistic VA and the consumer agent follow a very simple pricing rule, i.e., $p_i = c_i$. In addition, for any ranking, the two objective function values (total surplus and consumer surplus, respectively) are the same. It follows that the optimal rankings under the two objective functions are also the same. 
\begin{coro}{\textbf{\emph{(Altruistic/Consumer agent VA: Equilibrium and surplus allocation).}}}\label{cor:apca_pricing}\\ 
 In equilibrium, an altruistic VA and a consumer agent share the same optimal ranking and set all prices $p_i$ at cost $c_i$. In expectation, the consumer extracts the entire surplus.
% \begin{itemize}
%     \item[(i)]  For any given ranking $\bsigma$, the altruistic seller and the consumer agent set all prices $p_i$ at cost $c_i$. In expectation, the consumer extracts the entire surplus.
%     \item[(ii)]  In equilibrium, an altruistic seller and a consumer agent share the same optimal ranking and sets all prices $p_i$ at cost $c_i$. *** COMBINE THESE INTO ONE STATEMENT STARTING FROM (ii)*** 
% \end{itemize}
\end{coro}
In summary, the percentage of total surplus extracted by the profit-maximizing seller depends on the distribution of $\epsilon$ and is $50\%$ for the exponential distribution (we'll examine other distributions in Section \ref{sec:implication}). Under both altruistic and consumer agent VA, the seller's expected profit is zero. In addition, the equilibrium results (ranking, pricing, and consumer thresholds) are the same for a consumer agent and an altruistic VA, independent of the distribution of $\epsilon$. 

%% file: sec4_OptimalRanking.tex
ku\section{Optimal ranking}\label{sec:optimal_ranking}
Using Proposition \ref{prop:pp_eqm_general}, we can solve the optimal ranking problem by complete enumeration, i.e., by comparing the objective function values across all possible rankings. Finding the optimal ranking is then the combinatorial problem of searching among all $N!$ permutations, which becomes intractable for large $N$. In this Section, we provide two algorithms for efficiently computing or approximating the optimal ranking.

\subsection{The GPS algorithm} \label{sec_imp_alg}
 We propose an efficient algorithm which produces the optimal ranking based on checking local pairwise optimality. The algorithm, which we call the ``Greedy Pairwise Switch'' or ``GPS'' Algorithm, operates as  follows (we describe the algorithm for a profit-maximizing seller; the adjustments for other objective functions are straightforward). 
Let $\bsigma=(\sigma_1, \cdots, \sigma_N)$ be the current ranking and $\vR(\bsigma) $ be the set of all pairwise-switched rankings starting from $\bsigma$. For example, if $N=3$ and $\bsigma=(1,2,3)$, then $\vR(\bsigma) = \{(2,1,3), (1,3,2), (3,2,1)\}$. A pairwise-switched ranking is obtained by switching two products in the current ranking $\bsigma$ and keeping the ranks of all other products unchanged. Formally, 
\[    \vR(\bsigma) = \{ \bsigma' = (\sigma_1, \cdots, \sigma_{i-1},\sigma_j, \sigma_{i+1}, \cdots, \sigma_{j-1}, \sigma_i, \sigma_{j+1}, \cdots, \sigma_N )    , i<j     \} .    \]
Starting with $ \bsigma,$ the algorithm checks whether there is a local profit improvement for all $\bsigma' \in \vR(\bsigma)$. If there is no local improvement (i.e., $V^p(\bsigma)\geq V^p(\bsigma') \text{ for all } \bsigma'\in \vR(\bsigma)$), the algorithm returns $\bsigma$ as the optimal ranking. If there is a local improvement, the algorithm updates the current ranking to the local switch that achieves the largest improvement.
% \[ \bsigma \leftarrow  \argmax_{\bsigma'\in \vR(\bsigma)} V^p(\bsigma' )   \]
The algorithm terminates in finite time since we have a finite number of possible permutations and it never cycles since each switch provides a strictly positive improvement. 
% The algorithm is formally described in appendix.  

\begin{figure}[H]
    \centering 
    \subfloat[Complexity of the GPS algorithm.]{\includegraphics[width=0.45\textwidth]{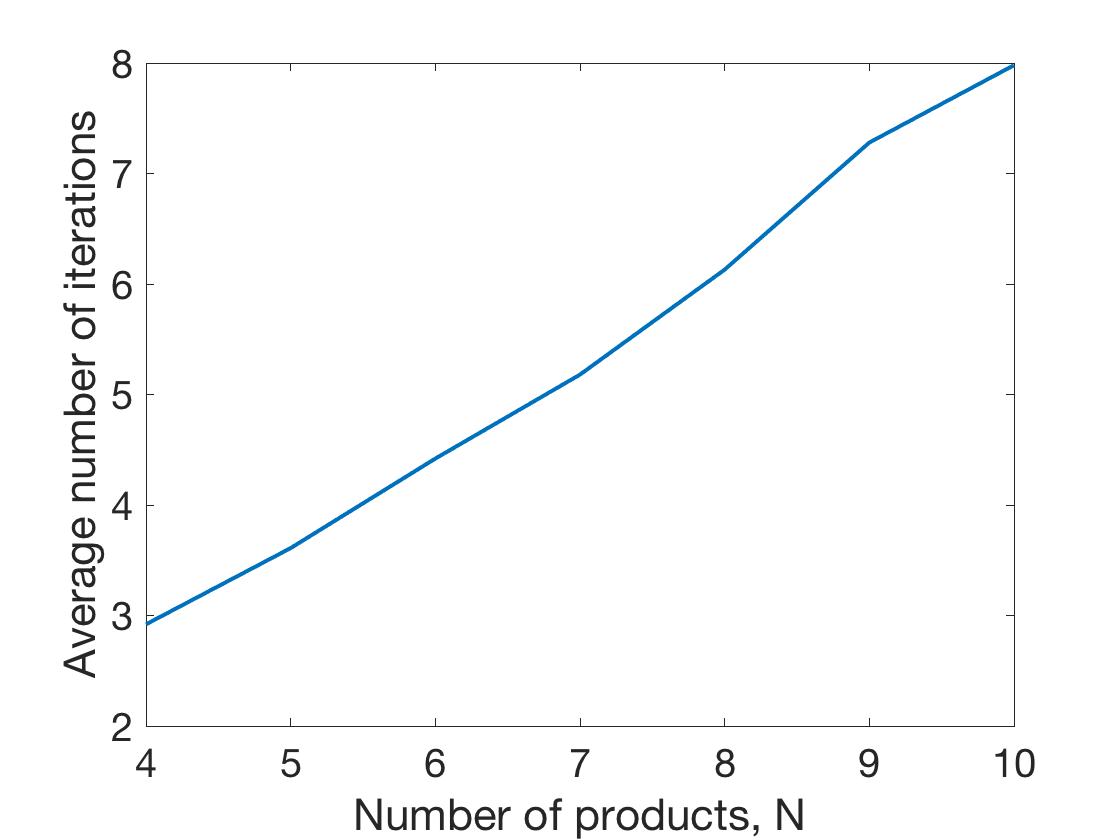}} 
    \subfloat[Complexity ratio for complete enumeration relative to the GPS algorithm.]{\includegraphics[width=0.45\textwidth]{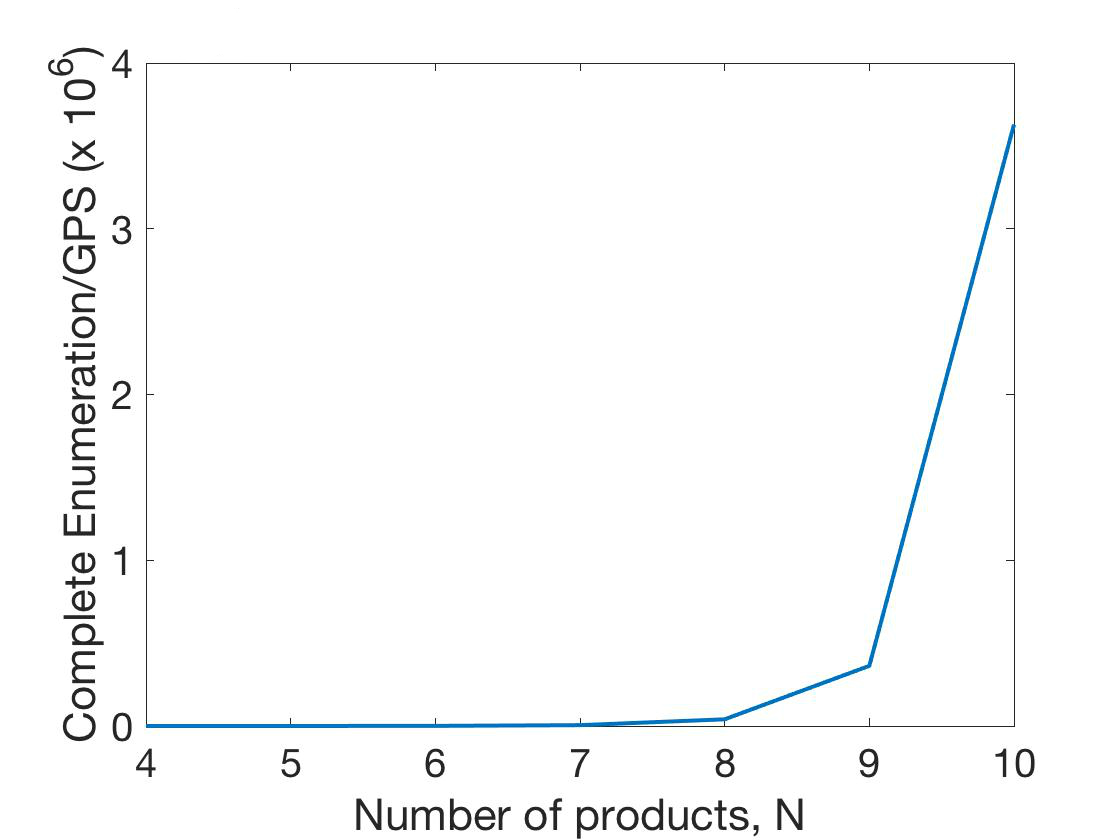}} 
    \caption{
    Complexity comparison between complete enumeration and the GPS algorithm. Shown is the average number of iterations as a function of the number of products, $N$, for each algorithm. The complexity of the GPS algorithm is linear in $N$ (left figure), a significant improvement over complete enumeration (right figure) that grows as $(N-1)!$.
    }
    \label{fig_pp_TimeComparison}
\end{figure}

Figure \ref{fig_pp_TimeComparison}(a) shows that the number of iterations required for the GPS algorithm increases almost linearly in $N$ and Figure \ref{fig_pp_TimeComparison}(b) shows the average iteration ratio between complete enumeration and the GPS algorithm. 
% The runtime for complete enumeration, which requires exhaustive search over $N!$ permutations increases dramatically. 
Clearly, GPS is much more efficient than complete enumeration, and in all the cases we examined in a large number of experiments, it achieved the optimal profit. 
%\commwb{?}
%*** CLEAN UP THE FIGURE AND THE DISCUSSION OF THE FIGURE; ALSO SHORTEN THE OBVIOUS DISCUSSION  *****

%*** Horizontal axis on both charts: Number of products, N Vertical axis on (a): Average number of iterations x 10 to the power 6 (instead of above the chart) (b): Complete enumeration/GPS ****

\subsection{Limiting cases and the double-rank approximation}\label{sec_limit_approx}
When the survival probability $\rho$ approaches $0$ (extremely impatient) or $1$ (extremely patient consumer), we obtain the following limits.

% *** The LaTex needs to be fixed below ****

\begin{prop}{\emph{\textbf{(Limiting case analysis)}}}\label{prop_ExtRho}\\
Assume the private valuations are exponentially distributed. Then, for each of the three objective functions (profit-maximizing, altruistic or consumer agent VA):
\begin{itemize}
    \item[(a)]  There exists a $\rho_0 > 0$ such that for all  $\rho \in [0, \rho_0)$, the descending ranking in value margins $(v_i + \EE(\epsilon)-c_i)$ is optimal. Further, for a profit-maximizing seller, the optimal prices are given by the monopoly prices  \ $p_i = c_i + \frac{1}{\alpha},~ i=1,2,...,N$.
    
    \item[(b)] There exists a $\rho_1 < 1$ such that for all $\rho \in (\rho_1, 1] $, the ascending ranking in value margins $(v_i + \EE(\epsilon)-c_i)$ is optimal.
\end{itemize}
\end{prop}
These limiting results suggest an approximating algorithm which we call the ``double-rank approximation.'' 

\begin{algrm}{\emph{\textbf{(Double-rank approximation).}}}
The seller compares the objective function values for \emph{only} the descending ranking and the ascending ranking in value margins $(v_i + \EE(\epsilon)-c_i)$, and selected the one that achieves the higher value.
\end{algrm}

% \subsection{Performance of the double-rank algorithm} \label{subsec_num_2rank}

The double-rank approximation is obviously computationally efficient since it only requires $ 2N $ iterations. But how close are its results to the optimal results?  We evaluate the efficiency of the algorithm through the ratio of expected profits under the double-rank algorithm to the optimal objective function value. Multiple numerical experiments show that in spite of the simplicity of the double-rank approximation, it is surprisingly efficient. 

\begin{figure}[H]
\centering
\subfloat[Gamma$(1, 1)$ = Exp(1)]{\includegraphics[width=0.33\textwidth]{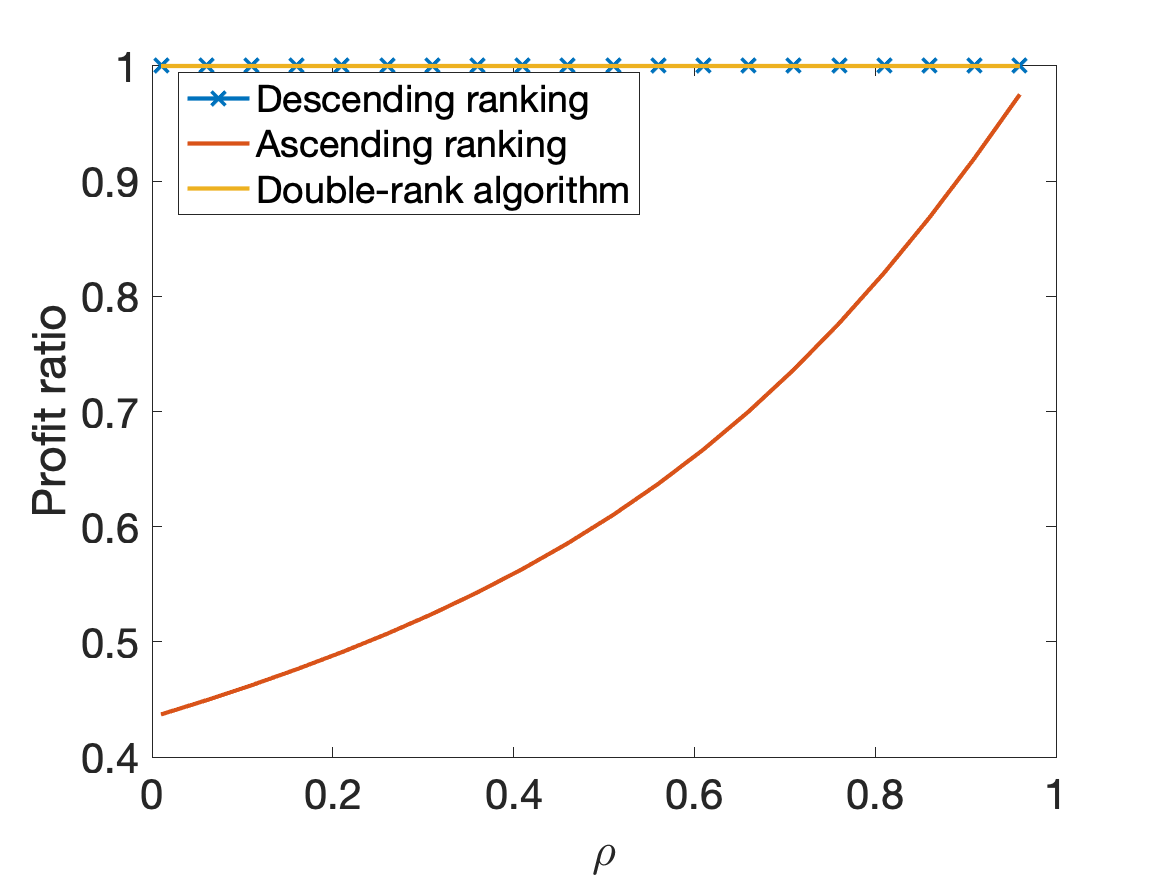}}
\subfloat[Gamma$(2, 0.5)$]{\includegraphics[width=0.33\textwidth]{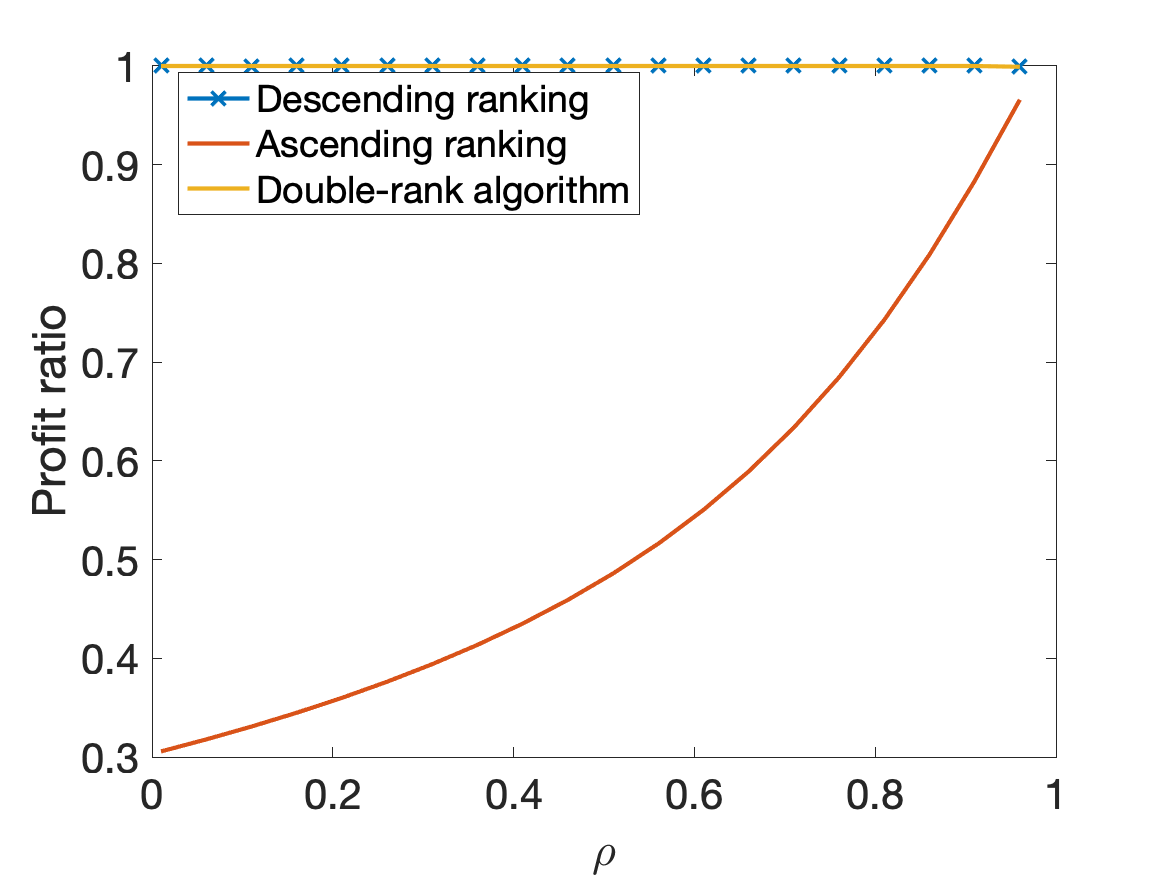}}
\subfloat[Gamma$(5, 10)$]{\includegraphics[width=0.33\textwidth]{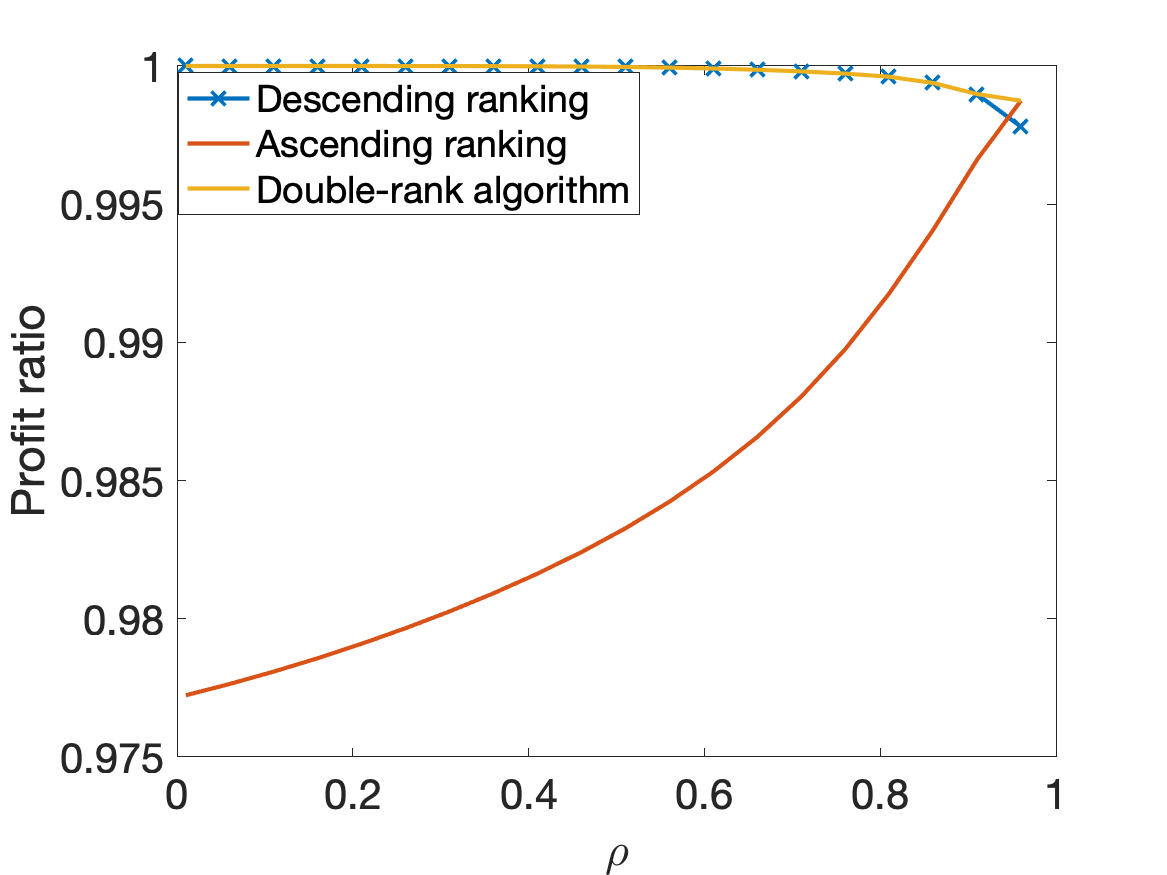}}
\caption{Efficiency of the double-rank approximation for Gamma-distributed private valuations. The figure shows the ratio of the expected profit achieved by the double-rank approximation to the optimal expected profit, and how it depends on the consumer patience level $\rho$. Shown are results for \textit{(i)} Exp(1), \textit{(ii)} Gamma (2,0.5), and \textit{(iii)} Gamma (5,10).}

% *** In the figure, change "surplus ratio" to "profit ratio.***

% The descending ranking remains optimal when the consumer patience is not close to $1 $ and the ascending ranking achieves optimal for extremely patient consumer. The double-rank algorithm achieves at least $95\%$ of optimal expected profit. }
\label{fig:pp_2rank}
\end{figure}

Our numerical experiments use the Gamma distribution, which is widely used in applications (Appendix \ref{app_numerical_setting} summarizes key features of the Gamma distribution and lists the parameters of our numerical experiments). We present here the results for three private valuation distributions: exponential with unit mean (Figure \ref{fig:pp_2rank}(a)), Gamma $(2, 0.5)$ (Figure \ref{fig:pp_2rank}(b)), and Gamma $(5, 10)$ (Figure \ref{fig:pp_2rank}(c)). As seen in the figures, the descending order remains optimal even for moderate values of $\rho$ and the ascending order kicks in only at $\rho = 0.99$ for the Gamma $(5, 0.1)$ and Exp$(1)$ cases, and at $\rho = 0.9$ when the distribution is Gamma $(5, 10)$ (where the private valuations have a higher variance $ab^2$). The results indicate that with lower information asymmetry, the descending order in value margins is more likely to be optimal. Throughout, the double-rank algorithm achieves at least $95\%$ of the optimal profit. Similar results are obtained for the other objective functions.
% ****CHECK*** \commwb{it is correct.}

% \commwb{I'll do this}
% *** PUT ALL THE PARAMETERS OF THE NUMERICAL EXPERIMENTS AND SALIENT FEATURES OF THE GAMMA DISTRIBUTION IN APPENDIX A and just refer to the Appendix later. *****   

%*** REVISE THE THREE FIGURES SO THE DESCENDING RANKING HAS A MARKER LIKE X so it will be separately visible in the charts. ***** \commwb{Fixed}

\noindent

% **** THIS DISCUSSION NEEDS TO BE CLEANED UP ****

% **** DISCUSS AND INTERPRET HERE CHARTS (i) and (ii) for the two shape parameters *** \commwb{in two charts we actually have the same shape parameter}

%% file: sec5_Implications.tex
\section{Implications}\label{sec:implication}

% *** CHANGE DISCUSSIONS BELOW BE REFERRING TO THE APPENDIX ********

In this Section, we discuss some of the implications of the foregoing results. We have considered three different objective functions for the VA: a consumer agent, an altruistic VA and a (seller) profit-maximizing VA. By proposition \ref{prop:apca_eqm}, under the first two objective functions (consumer agent or altruistic VA), $p_i = c_i$ for all $i$ and the consumer extracts the entire surplus. When the VA maximizes the seller's profit, the optimal prices and surplus allocation depend on the consumer patience parameter and on the distribution of private information. When the private information is exponentially distributed, the surplus is equally split between the consumer and the seller, independent of the other problem parameters. The results are more complex when the private information follows a general distribution and we illustrate them using numerical examples.  

We consider $N=6$ products with expected valuations $(0, 1/6, 1/3, 1/2, 2/3, 5/6)$ and zero costs $(c_i = 0)$ (see Appendix \ref{app_numerical_setting} for a full listing of the parameters). Fixing these parameters, we first examine the structure of the optimal solution (pricing and ranking) when the private valuations are exponentially distributed (Subsection \ref{subsusec:va_rho_pricing}). We then consider the structure of the solution when the private valuations follow the more general gamma distribution (Subsection \ref{subsubsec:va_rho_ranking}) and we finally consider how information asymmetry affects the pricing, ranking and surplus allocation (Subsection \ref{subsec:va_info}).

% We briefly discuss the numerical setting here. The results in section \ref{subsec:diff_obj} hold independent of the distribution of private information $\epsilon$. In section \ref{subsec:va_rho}, we first use exponential distribution to illustrate the effect of consumer patient parameter $\rho$ on the optimal ranking and optimal pricing (section \ref{subsusec:va_rho_pricing}). In the non-extreme region of consumer patience parameter $\rho$, the optimal ranking can be very complicated and we display one example using Gamma distribution in section \ref{subsubsec:va_rho_ranking}. 
% \subsection{Structure of the profit-maximizing solution} \label{subsec:va_rho}
    % In this subsection, we study the structure of the profit-maximizing solution and how it depends on the consumer patience parameter $\rho$. We first consider the case where the private valuations are exponentially distributed, i.e., $\epsilon \sim \exp(1)$, and then the case where they follow a gamma distribution. 
    
    \subsection{Ranking and pricing: exponential case} \label{subsusec:va_rho_pricing}
\begin{figure}[H]
        \centering
        \subfloat{\includegraphics[width=0.5\textwidth]{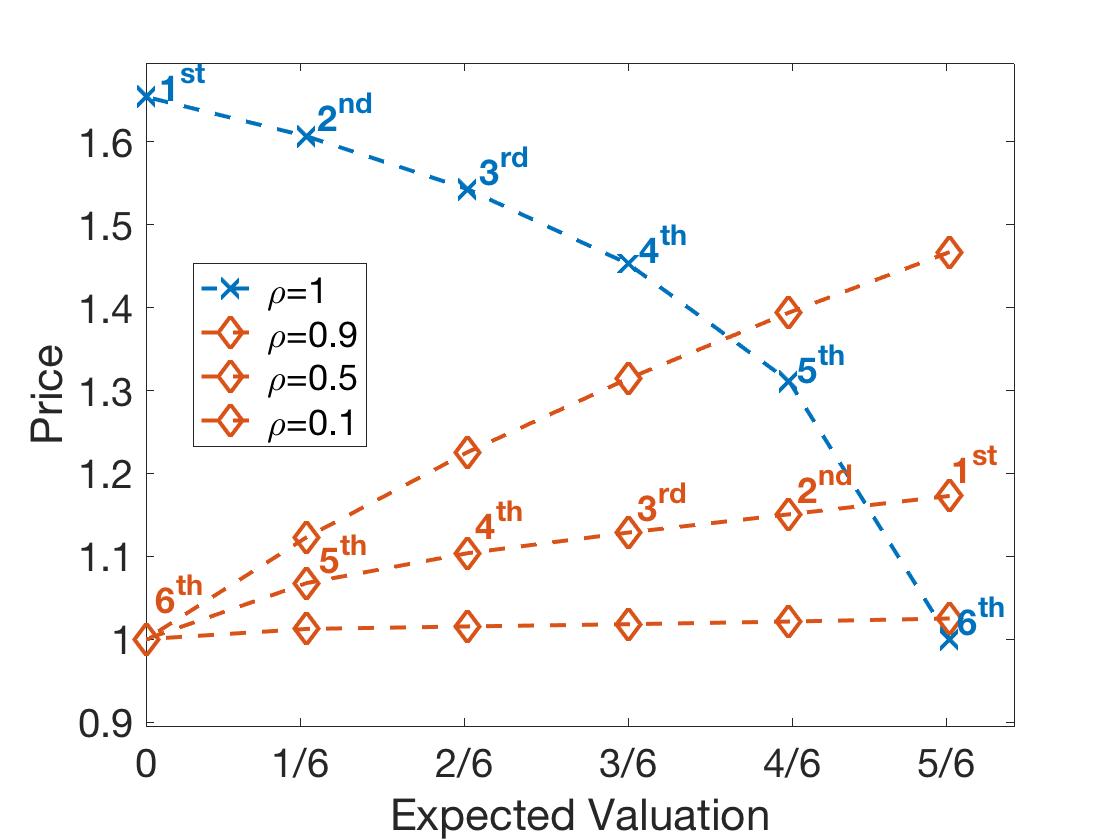}}
        \caption{Comparison of virtual assistant prices when $\epsilon$ is exponentially distributed, $\epsilon \sim \exp(1)$, for $\rho=0.1, 0.5, 0.9$ (brown lines) and $\rho=1$ (blue line). The horizontal axis shows the six product valuations.
        % *** SHOW THE NUMBERS ON THE AXIS AS 1/6, 1/3, 1/2 etc. ****. 
        The vertical axis shows the corresponding prices. The presentation rank is either descending (brown lines) or ascending (blue line) in the expected valuations.}
        \label{fig:va_price_patience}
    \end{figure}

    We first consider the problem with $\epsilon \sim \exp(1)$ for different levels of the consumer patience parameter $\rho$ (Figure \ref{fig:va_price_patience}).

    % 1. explain the ranking in the graph. (explain the reason).
    % 2. effects of ranking on the optimal prices
    % 3. markup as a function of other product.
When $\rho$ is 0.9 or less (brown lines), the optimal ranking is \textit{descending} in the expected valuations, and the seller ranks the most valuable product (with expected valuation $5/6$) first, similar to the limiting results in Proposition \ref{prop_ExtRho}(a). When $\rho=1$ (blue line), the optimal ranking is \textit{ascending} in the expected valuations, and the seller ranks the most valuable product last, as shown in Proposition \ref{prop_ExtRho} (b).
    
One might expect higher-valued products to command higher prices. In this case, however, Corollary \ref{cor:prop_pp_eqm_exp} shows that for a given ranking, the optimal price of each product $i$ is \textit{independent} of its own expected valuation, and it depends primarily on its rank which determines $\rho V_{i+1}^p$ through the expected valuations of products $i+1,i+2, \ldots N$. 
The pricing decision balances two considerations: on the one hand, an impatient consumer is unlikely to buy the products presented late, suggesting a higher margin for products presented earlier, which are more likely to be bought by the consumer. On the other hand, the different products effectively compete with one another, with less candidates remaining to be viewed, the seller has greater market power so it can extract higher margins (at the extreme, when only one product is left, Corollary \ref{cor:prop_pp_eqm_exp} shows that the seller charges the monopoly price). The optimal balance depends on the consumer's impatience parameter $\rho$: When the consumer is relatively impatient, the seller prefers to charge higher margins on the products presented earlier since there is only a small probability that she will buy products presented later. When the consumer is patient, there is a high probability that the consumer will wait and the seller charges higher margins on the products presented later. Accordingly, Figure \ref{fig:va_price_patience} shows that when $\rho$ is 0.9 or less (brown lines), the margins decrease as the product rank increases, whereas for $\rho=1$ (blue line), the margins increase along with the product rank.    

% ***** Effects on rho for fixed ranking *****

Figure \ref{fig:va_price_patience} also shows that for any given ranking $\bsigma$, each product's optimal price increases in the consumer patience parameter $\rho$. The Proposition below generalizes this finding.  
\begin{prop}{\emph{\textbf{(Monotonicity in Optimal Prices)}}}\label{prop_MonoPrice}\\ 
When the private valuations $\epsilon$ are exponentially distributed, under any given ranking, the optimal prices increase in the consumer patience parameter $\rho$.
\end{prop}
Proposition \ref{prop_MonoPrice} follows from the fact that under any given ranking, product $i$'s optimal price is the sum of the monopoly price $\frac{1}{\alpha}+c_i$ and the markup $\rho V^{\sigma_i+1}$, which in turn is increasing in $\rho$.  Intuitively, a more patient consumer receives a higher surplus, which allows the seller to extract more of that surplus through higher prices. 

    \subsection{Ranking and pricing: Gamma private valuations}
    \label{subsubsec:va_rho_ranking}
    When the private valuations are not exponentially distributed, the optimal solution becomes more complex. By Proposition \ref{prop_ExtRho}, at the limits as $\rho$ goes to zero or $1$, the optimal rankings are monotone in the products' value margins. In the example below, we show that, for the Gamma distribution, \emph{(i)} monotone rankings remain optimal for a wide range of (although not all) $\rho$ values; and \emph{(ii)} in the narrow range where the optimal ranking is not monotone, it changes quickly from descending to ascending through multiple switches. 

\begin{figure}[H]
        \centering
        \subfloat{\includegraphics[width=0.6\textwidth]{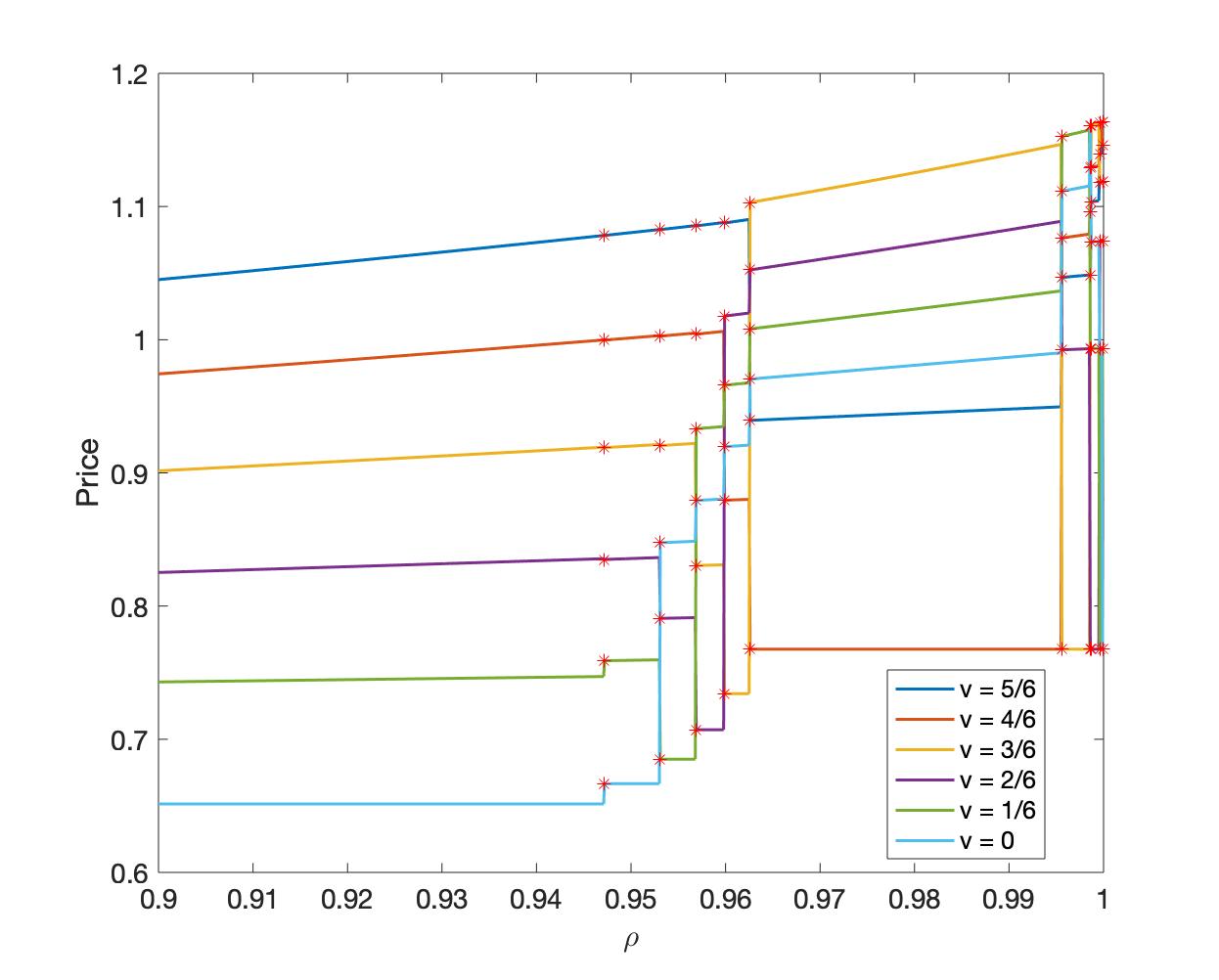}}
        \caption{\emph{Optimal prices for different values of consumer patience parameter $\rho$ when $\epsilon \sim $ Gamma(2, 0.5).} Shown are the optimal prices for the six products in each range of $\rho$ values in $[0.9,1]$ (the behavior in $(0,0.9)$ is similar to the behavior in $(0.9, 0.9478)$). The red stars identify changes in the optimal ranking.}
        \label{fig:gamma_price_va}
    \end{figure}

   We consider $\epsilon \sim \text{Gamma}(2, 0.5)$ and vary $\rho$ from zero to $1$ (Table \ref{tb:ranking_switch} and Figure \ref{fig:gamma_price_va}). As shown in the Table, the descending ranking is optimal for all $\rho \in [0, 0.94718)$, and the ascending ranking is optimal when $\rho \in (0.99988, 1]$). In-between ($\rho \in (0.94719, 0.99988)$), the solution (ranking and product prices) changes ten times.
   
        \begin{table}[H]
        \centering
        \begin{tabular}{||c|| c |c | c| c|c|c||}
         \hline
        $i$  & $1$ &   $2$& $3$ &  $4$ &$5$ &$6$\\ 
         \hline
        Expected  Valuation &$0$  & $\frac{1}{6}$ &   $\frac{1}{3}$& $\frac{1}{2}$ &  $\frac{2}{3}$ &$\frac{5}{6}$ \\ [0.5ex] 
         \hline\hline
         $(0.00000, 0.94718] $ &6 &5 &4 &3 &2 &1 \\ 
         \hline
         $(0.94719, 0.95305] $ &6 &5 &4 &3 &1 &2\\
         \hline
         $(0.95306, 0.95687] $&6 &5 &4 &2 &1 &3\\
         \hline
        $(0.95688, 0.95984] $ &6 &5 &3 &2 &1 &4\\
         \hline
         $(0.95985, 0.96250] $ &6 &4 &3 &2 &1 &5\\ 
        \hline
        $(0.96251, 0.99552] $ &5 &4 &3 &2 &1 &6\\
        \hline
        $(0.99553, 0.99859] $&4 &3 &2 &1 &5 &6 \\\hline
        $(0.99860, 0.99865] $&3 &2&4 &1 &5 &6 \\\hline
        $(0.99866, 0.99950] $ &3 &2 &1 &4 &5 &6 \\\hline
        $(0.99951, 0.99956] $ &2 &3 &1 &4 &5 &6 \\\hline
        $(0.99957, 0.99987] $ &2 &1 &3 &4 &5 &6 \\\hline
        $(0.99988, 1.00000] $ &1 &2 &3 &4 &5 &6 \\[1ex] 
         \hline
        \end{tabular}
        \caption{\emph{Optimal rankings for different consumer patience levels $\rho$} when $\epsilon \sim \text{Gamma}(2, 0.5)$. Shown are the optimal rankings for each range of $\rho$ values in (0,1]. For example, the optimal ranking permutation is $(6, 5, 4, 3, 1, 2)$ when $\rho \in (0.94719, 0.95305]$.} 
        \label{tb:ranking_switch}
    \end{table}
    
    Similar to the exponential case, the consumer patience parameter $\rho$ influences the optimal prices through the markup and the optimal ranking. As we observe in Figure \ref{fig:gamma_price_va}, in each interval where the optimal ranking remains constant, the optimal price is increasing in $\rho$ for each product. 

\subsection{Effects of information asymmetry} \label{subsec:va_info}
In this Subsection, we study the effects of information asymmetry on the allocation of surplus.
% To understand this, we introduce the notion of the consumer's private information and the seller's information. 
Whereas the seller knows only the $v_i$s, the consumer can also learn the realization of the $\epsilon_i$s. This gives the consumer an informational advantage that, intuitively, should increase with the variance of $\epsilon$. To study how this information asymmetry affects our results, we consider Gamma-distributed private valuations with shape parameter $a$ and scale parameter $b$, fixing the mean at $\EE(\epsilon) = ab = 1$ and changing the shape parameter $a$. With a constant mean, increasing the shape parameter $a$ is equivalent to reducing the variance ($\Var(\epsilon)=ab^2= \frac{1}{a}$) of the consumer's private information.

\begin{figure}[H]
        \centering
    `    \includegraphics[width= 0.6\textwidth]{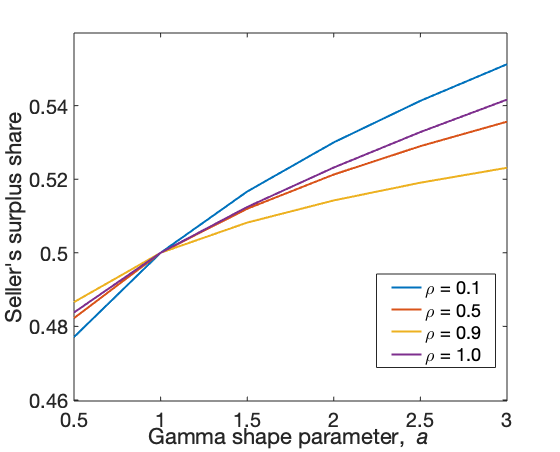}
        \caption{Seller's surplus share for a profit-maximizing VA. The private information $ \epsilon \sim \emph{Gamma}(a, b), ab = 1$. As $a$ increases, the variance $1 \over a$ decreases.}
        \label{fig:va_info_surplusallocation}
        \end{figure}

     Figure \ref{fig:va_info_surplusallocation} shows that as the shape parameter $a$ increases, the seller's surplus share increases as expected. When $a=1$, the distribution of $\epsilon$ is exponential and the seller extracts $50 \%$ of the surplus. The limit as $a \rightarrow \infty$ (not shown) is deterministic with no private information.  In this case, the seller extracts all the surplus and consumer gets $0.$
    %  ****Figure vertical axis: Seller's surplus share.
    %  Horizontal axis: Gamma shape parameter, a. ********** \commwb{italic a}
     
    %  **** JUST EXPLAIN THE LIMITS AND SKIP THE EFFECT OF RHO****  
    
%   In addition, as the consumer patience level $\rho$ increases, the slope for increment decreases, implying that it's more profitable for a seller to invest in obtaining information when the consumer patience level is low.  *** WHY DOES THIS HAPPEN ???? ***** \commwb{It's hard to provide analytical explanation ... Shall we remove this  sentence?}

%% file: sec5_webInterfaceComparison.tex
\section{Web interface vs. virtual assistant} \label{sec:web_vs_va}
\vspace{20pt}
% *** This section requires a more comprehensive revision following our discussion ***
% *** Start with the motivation for the comparison.  Move the discussion of commitment power to the pricing comparison, later **** 

% **** THIS SECTION WILL NEED TO BE CLEANED UP AND COMPLETED ONCE WE ARE DONE WITH APPENDIX C *****

Traditional web-based sellers such as Amazon, eBay and Taobao dominate today's electronic commerce market. What would be the effect of the predicted shift to sales through virtual assistants on pricing, seller profits and consumer surplus? To answer this question, we model sales through a web interface and compare the results to those we obtained for a virtual assistant. We focus on a (seller) profit-maximizing VA.
% in section \ref{subsec:web-profit-max} and then analyze what happens under alternative objective functions in section \ref{subsec:web-other-obj}.
 
\subsection{Web Interface: Model}\label{subsec: WI Model}

Consider a web interface that enables the presentation of $k$ products per page. In our model, the consumer specifies what she is looking for and the seller presents the available products a page ($k$ products) at a time. The consumer then examines the entire page and decides whether to choose a product within that page, or to proceed to the next page. The process continues until the consumer exits.

As in our virtual assistant model, the consumer is impatient and stays on for an exponential amount of time with mean $1 \over \tau_c$. We assume that with the web interface, the time to evaluate a page is exponentially distributed with rate $\kappa \tau_p / k$, where $\kappa$ is an acceleration factor specifying how fast the consumer views one page ($1 \le \kappa \le k$). When $ \kappa = k$, the expected time to view an entire page is the same as that for evaluating one product with the virtual assistant, whereas when $\kappa =1$, viewing a page takes $k$ times as long as evaluating a single product with the virtual assistant.  We denote by $u_{l,i}=v_{l,i}+\epsilon_{l,i}-p_{l,i}$ the realized utility when the consumer buys the $i^{th}$ product on page $l$, and by $v_{l,i}$ the valuation of the $i^{th}$ product on page $l$. 

With this specification, the virtual assistant corresponds to the special case of a web interface with $k=\kappa=1$. We compare the behavior of prices and the allocation of surplus between the web interface and the virtual assistant.
% *** As what?*** \commwb{as when it's  divisible.}.

% *** What happens when N is not divisible by k? Will it change the results for the last page and leave the recursions intact? ****** \commwb{Yes? The analysis for the remaining pages (except for the last page) remains the same, because the effect from the last page is fully captured in the $\delta$ term}  *** Assuming that's the case, we should say something like: For simplicity, we assume in what follows that $N$ is divisible by $k$. Otherwise, the last page will display less than $k$ products and the optimization problem remains the same.

% Taking into account the possibility that the consumer will exit (rate $\tau_c$), the consumer's expected dwell time on a page is $\frac{1}{\kappa \tau_p/k+\tau_c}$. The corresponding impatience parameter $\rho_{web}$ (defined as the probability that ****) is thus \[\rho_{web}=\frac{\kappa \tau_p/k}{\kappa \tau_p/k+\tau_c}.\] When $\kappa=1$, the consumer spends on average the same amount of time viewing one product under the two interfaces. When $\kappa =k$, viewing one full page under the web interface takes on average the same amount of time as viewing one product using the virtual assistant. With this specification, the virtual assistant corresponds to the special case $k=1$, $\kappa=1$.

\subsection{Profit-maximization}\label{subsec:web-profit-max}
     \subsubsection{Optimal pricing}
    For a given ranking, the equilibrium prices and the consumer's optimal thresholds under the web interface are given by Proposition \ref{prop:pp_eqm_web_general}, whose straightforward proof is omitted.
    %are calculated by backward induction (the proof is in Appendix \ref{prof:pp_eqm_web_general}). Here $u_{l,i}=v_{l,i}+\epsilon_{l,i}-p_{l,i}$, is the realized utility that the consumer accepts $i^{th}$ product at page $l$ and $v_{l,i}$ is the valuation of the $i^{th}$ product on page $l$. 
    
    % \commwb{Note that in the case when mod(N, k) $\neq 0,$ the total number of pages is $\lceil\frac{N}{k} \rceil $ and the entire analysis is the same.}
% In general, the optimal pricing problem can be solved by backward induction and the optimal prices solve a system of equations induced by the first order condition. We complete our analysis on web interface seller for profit maximizing seller by the following backward induction. We provide the following proposition for web-interface optimal pricing under fixed order.\\

\begin{prop}{\emph{\textbf{(Web-interface equilibrium for a given ranking with  $\epsilon \sim F(\cdot)$)}}}\label{prop:pp_eqm_web_general}\\
Given $\bsigma= (1, 2, \cdots, N)$, an equilibrium $(\bdelta, \vp)$  must satisfy the following recursion:
\begin{enumerate}[label=(\alph*)]    
    \item In stage $n : = \lceil\frac{N}{k} \rceil $,\vspace{-2mm}
    \begin{itemize}
        \item[] The consumer's threshold is $\delta_n = 0$.
        \item[] For any price vector $\bm{p}$, the probability of purchasing product $i$ in stage $n$ is given by
        \[\PP_{n,i}=\PP( u_{n,i} \geq \delta_n, u_{n,i} \geq u_{n,j} \text{~for all~} j \neq i, \text { $j$ is on page $n$}), \]
        where $u_{n,i}=v_{n,i}+\epsilon_{n,i}-p_{i}$. 
        \item[] The seller's expected profit in stage $n$ is given by
        \[V^p_n   =  \max_{\bm{p}\in\mathbb{R}^{k}}  \sum_i \PP_{n,i}(p_{n,i}-c_{n,i}).\]
        Let $\bm{p}_n$ be the maximizer of the above equation, then the consumer's expected surplus in stage $n$ is given by  
         \[V^c_n  =   \sum_i \PP_{n,i}\EE(\epsilon_{n,i}+v_{n,i}-p_{n,i}|u_{n,i} \geq \delta_n, u_{n,i} \geq u_{n,j}, \text{~for all~} j \neq i, \text { $j$ is on page $n$}).\]
    \end{itemize}
    \item In stage $l= n - 1, \cdots, 1, $\vspace{-3mm} 
        \begin{itemize}
        \item[] The consumer's threshold is $\delta_{l}  = \rho V^c_{l+1}$.
        \item[] For any price vector $\bm{p}$, the probability of purchasing product $i$ in stage $n$ is given by
        \[\PP_{l,i} = \PP( u_{l,i} \geq \delta_{l}, u_{l,i} \geq u_{l,j}, \text{~for all~} j \neq i, \text { $j$ is on page $l$}),   \]
        where $u_{l,i}=v_{l,i}+\epsilon_{l,i}-p_{i}.$
        \item[] The seller's expected profit in stage $l$ is given by
        \[V^p_{l}   = \max_{\bm{p}\in\mathbb{R}^{k}}  \sum_i \PP_{l,i} (p_{l,i}-c_{l,i})+  \rho (1-\sum_i\PP_{l,i})  V_{l+1}^p. \]
        Let $\bm{p}_{l}$ be the maximizer of the above equation, then the consumer's expected surplus in stage $l$ is given by  
         \[V^c_{l} =  \sum_i \PP_{l,i} \EE(\epsilon_{l,i} + v_{l,i}-c_{l,i}|u_{l,i} \geq \delta_{l}, u_{l,i} \geq u_{l,j}, \text{~for all~} j \neq i, j\text{ is on page}l )  + (1-\sum_i \PP_{l,i}) \rho V_{l+1}^c .\]
        \end{itemize}
\end{enumerate}
\end{prop}
     Proposition \ref{prop:pp_eqm_web_general} holds regardless of whether $N$ is divisible by $k$.
     %, the number of pages is $\lceil\frac{N}{k} \rceil$ and the optimization equations remains the same whether it is divisible or not. 
     It enables us to compute the optimal ranking following the methodology of Section \ref{sec:implication}. 
    
\begin{figure}[H]
    \centering
    \subfloat[$\rho= 0.1$]{\includegraphics[width=0.45\textwidth]{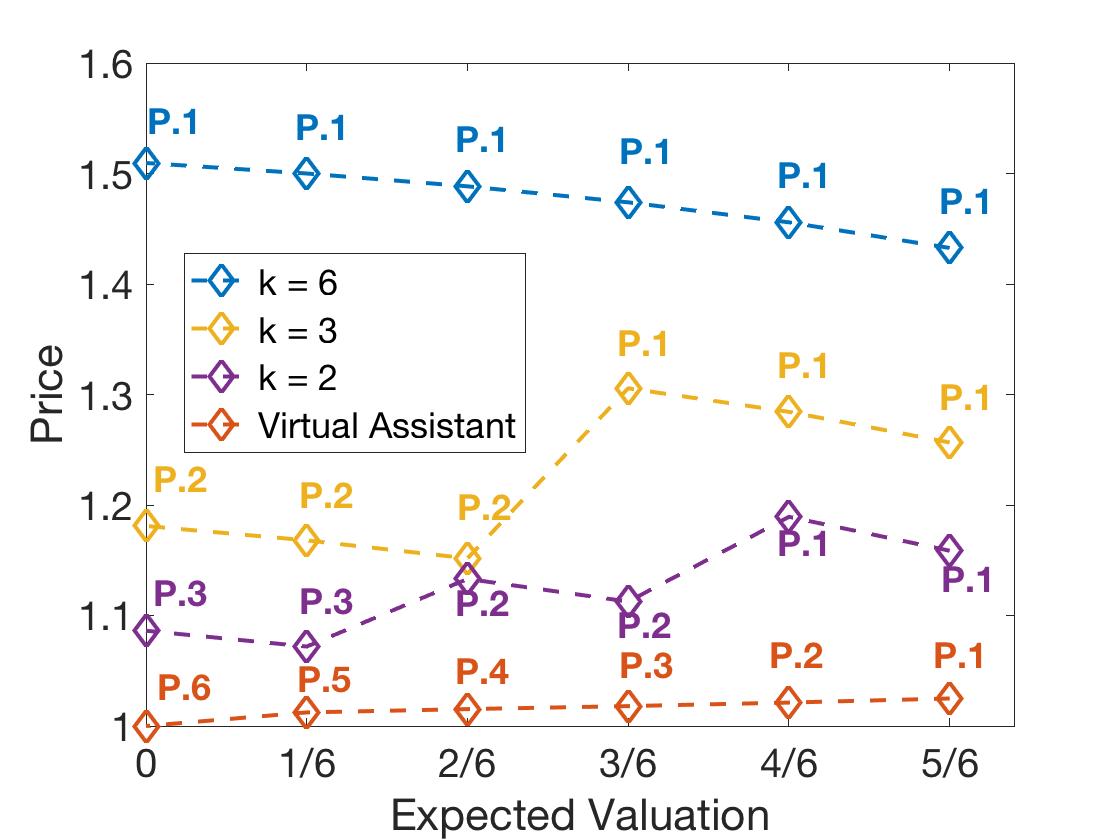}}
    \subfloat[$\rho= 0.9$]{\includegraphics[width=0.45\textwidth]{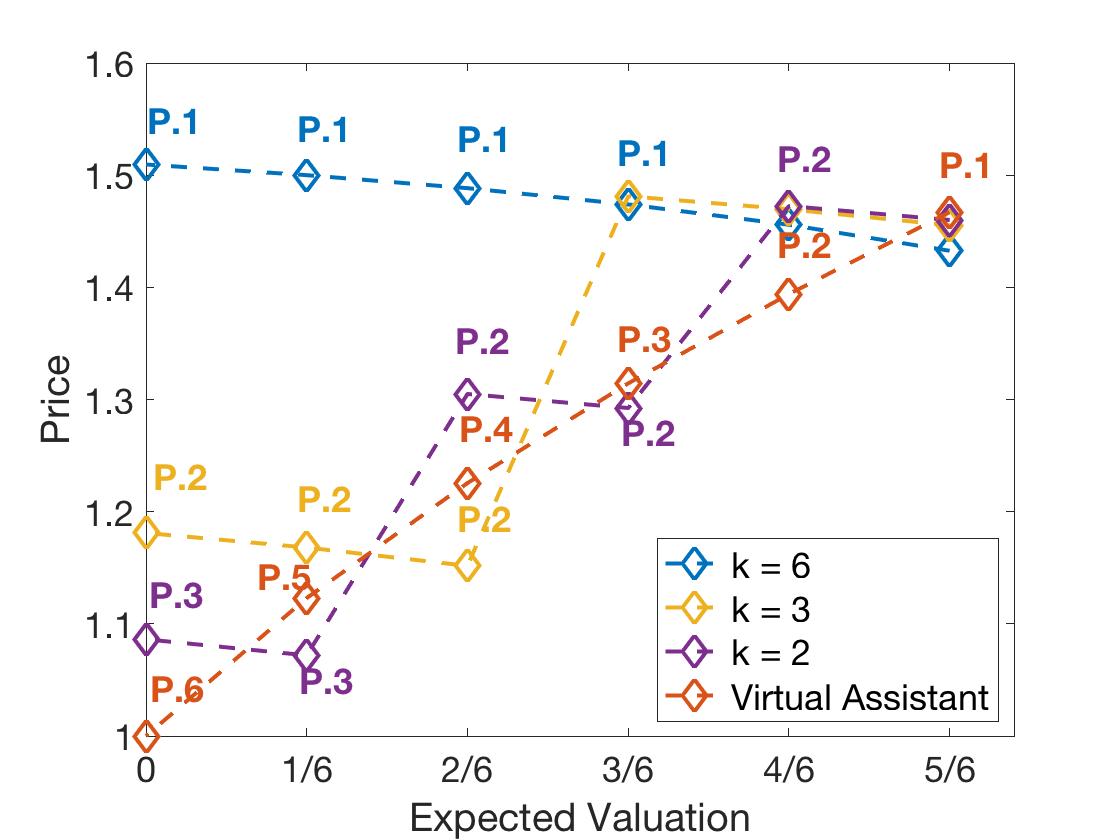}}\\
    \subfloat[$\rho=1$]{\includegraphics[width=0.45\textwidth]{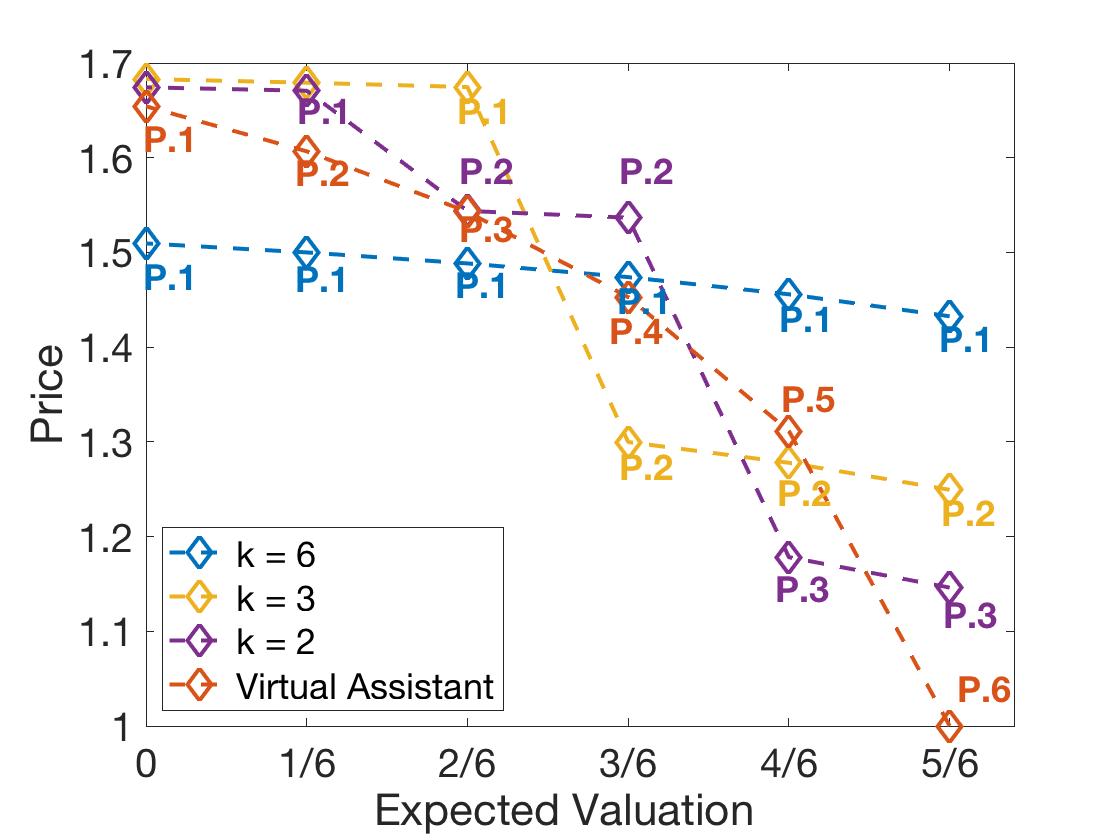}}
    \caption{Price comparison between virtual assistant and web interface with $k=2,3$ or $6$ products per page, with $\kappa=1$. We present the optimal prices under the optimal ranking for $\rho=\frac{\tau_p}{\tau_c+\tau_p}=0.1$ (a) , $0.9$ (b), and $1$ (c). In the figure, the label $P.i$ means that the product is presented on page $i$. Some page labels are not shown for nearly-overlapping data points.}
    \label{price_web_sim1}
    \end{figure}

    We next illustrate the effects of the user interface on price behavior using numerical examples. As before, we first consider a seller selling $6$ products with expected valuations $0, 1/6, 1/3, 1/2, 2/3$  and $5/6$, where the private valuations are exponentially distributed with unit mean, $\epsilon \sim \exp(1)$. We compare the virtual assistant and the web interface with $k=2,3$ or $6$ products per page in Figure \ref{price_web_sim1}. 
    
  \textit{Across pages}, the price pattern we observe is similar to the one we obtained for the virtual assistant. For low and moderate values of $\rho$, the virtual assistant ranked products in \emph{descending} order of product valuations and the prices were \emph{increasing} in product valuations. Similarly, the web-interface places the products with higher valuations on earlier pages (Figure \ref{price_web_sim1}(a),(b)), and these products are priced higher. Although under the web-interface, the there is no elegant decomposition of price into a monopoly price and a continuation markup, the intuition carries over: when the consumer is willing to accept a product presented early, the seller infers that she has a favorable private valuation, which the seller exploits to extract a higher profit. For large $\rho$, the optimal ranking and pricing rules for the virtual assistant were reversed: the optimal order was \emph{ascending}, and prices were \emph{descending}, in product valuations. We observe a similar pattern under the web-interface: the order is descending in the product valuations and the optimal prices are higher for the earlier pages (Figure \ref{price_web_sim1}(c)). 

  However, the prices \textit{within} a page are monotone \textit{decreasing} in the product valuations in all three figures. We explain in detail why this happens in Appendix \ref{app_twoproduct} for the case of two products on a single page. In the Appendix, we prove that for a web interface with two products on the same page, the optimal prices reverse the order of product valuations. This happens because the equilibrium price of $p_1$ is not a function of $v_1$, but is an increasing function of the other product's price margin $v_2-p_2$. Similarly, the equilibrium price of $p_2$ is not a function of $v_2$, but is an increasing function of the other product's price margin $v_1-p_1$. An increase in $v_1$ does not have a direct effect on product $1$'s equilibrium price, but it increases $p_2$ through the increase in the price margin of product $1$ (equation (\ref{focp2})  in Appendix \ref{app_twoproduct}). However, increasing on $p_2$ further decreases the value-price margin of product $2$, hence it further decreases the equilibrium price of product $1$. And we have the prices are in reverse order of valuations.

    %    start from (a), for fix rho;
    %    start from (b), for fix rho;

    %   Figure *** shows that contrary to one's intuition, the prices \textit{within} each page are monotonically \textit{decreasing} in the product valuations: products with higher expected valuations are priced lower within each page.  In Appendix **** we show why this happens with $k=2$ products per page.  

    \subsubsection{Surplus allocation}
\begin{figure}[H]
    \centering
    \subfloat[$\rho=0.1, k=2.$ ]{\includegraphics[width=0.32\textwidth]{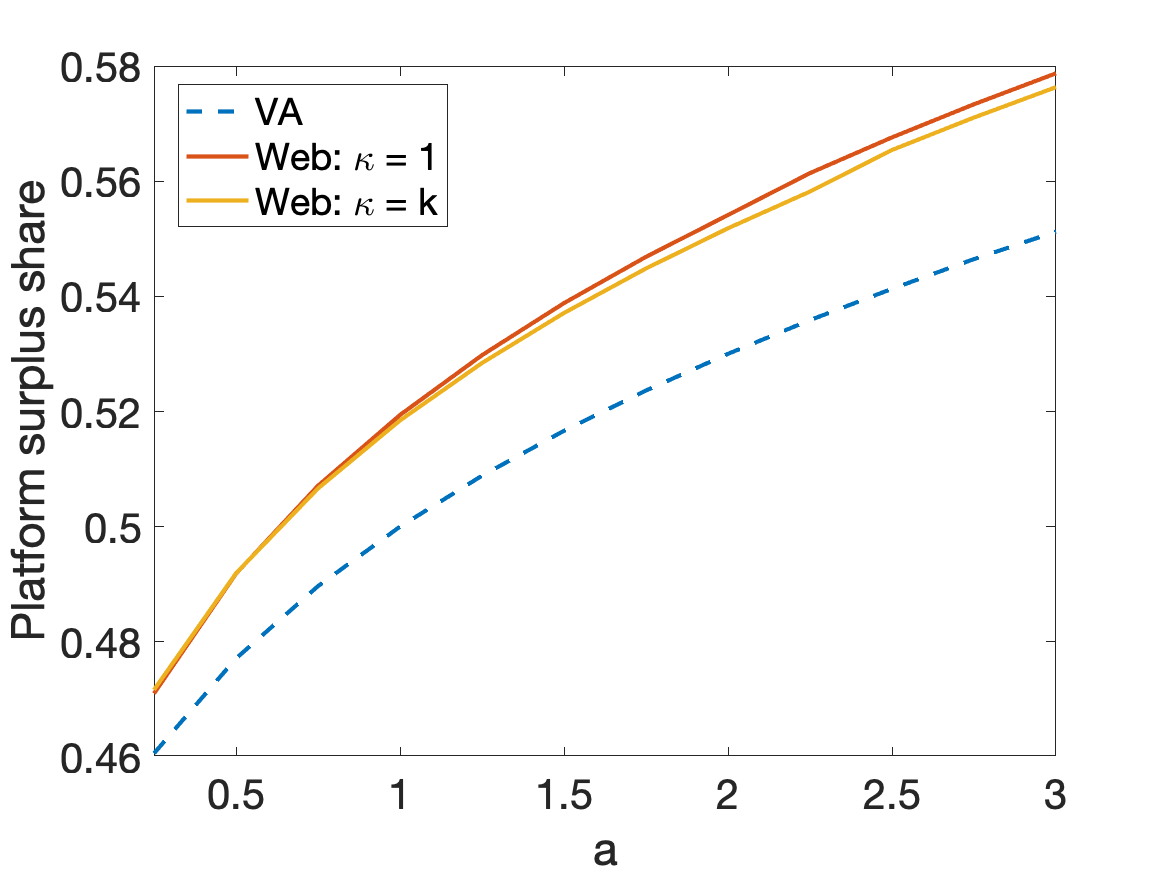}}
     \subfloat[$\rho=0.1, k=3.$ ]{\includegraphics[width=0.32\textwidth]{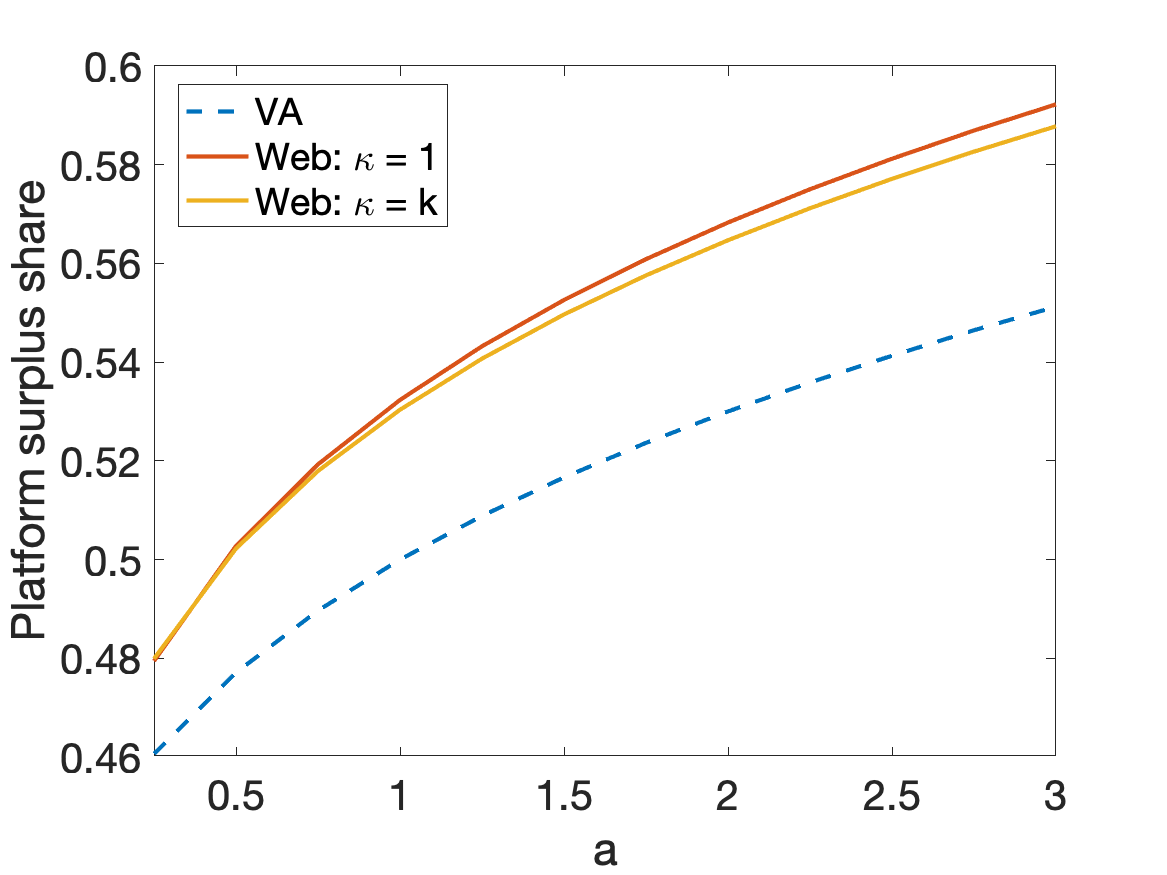}}
    \subfloat[$\rho=0.1, k=6.$ ]{\includegraphics[width=0.32\textwidth]{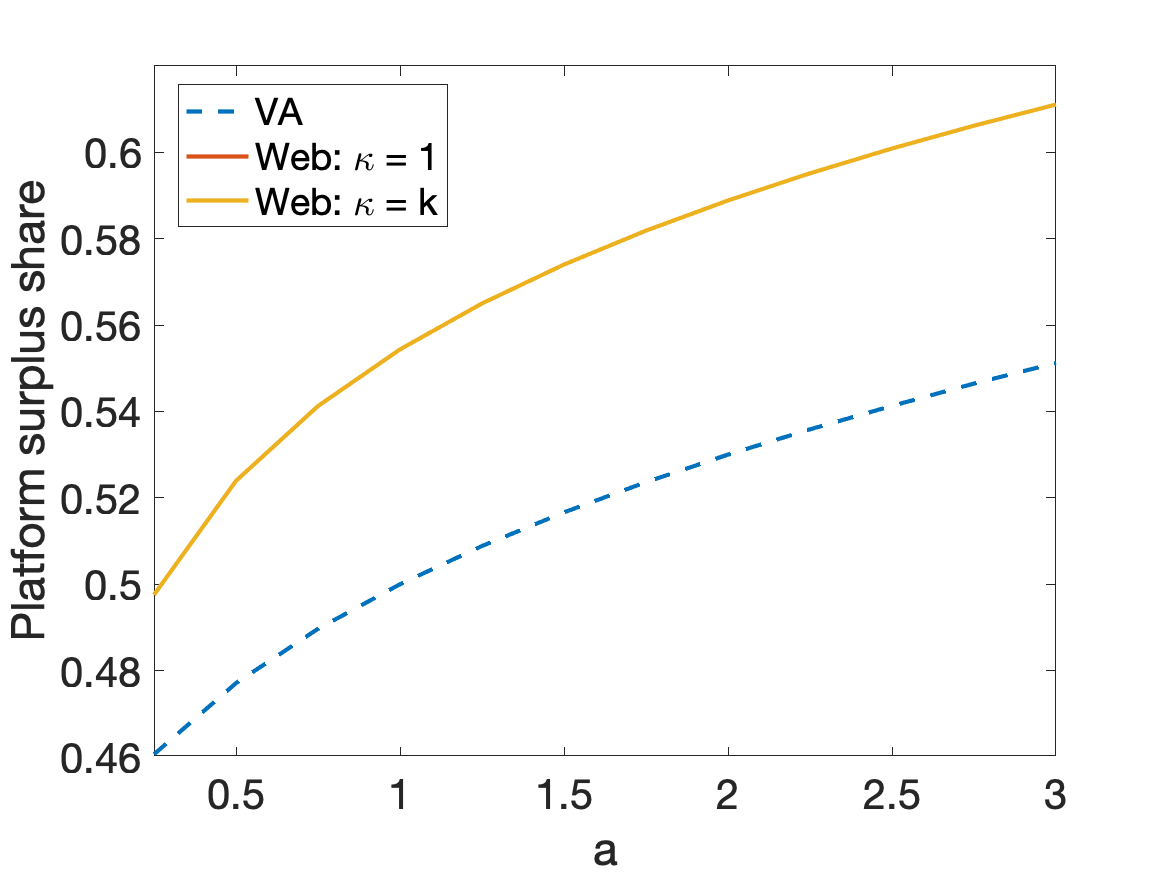}}\\
    \subfloat[$\rho=1, k=2.$ ]{\includegraphics[width=0.32\textwidth]{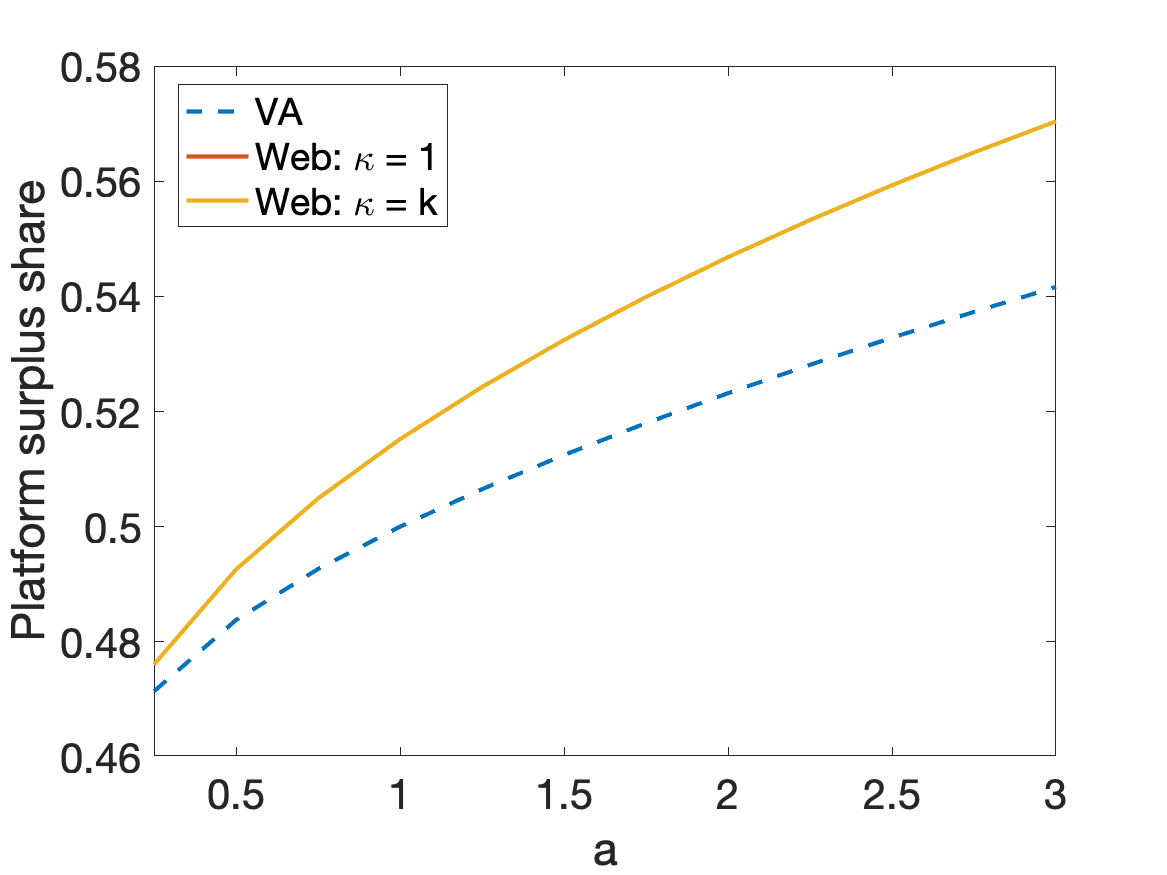}}
    \subfloat[$\rho=1, k=3.$ ]{\includegraphics[width=0.32\textwidth]{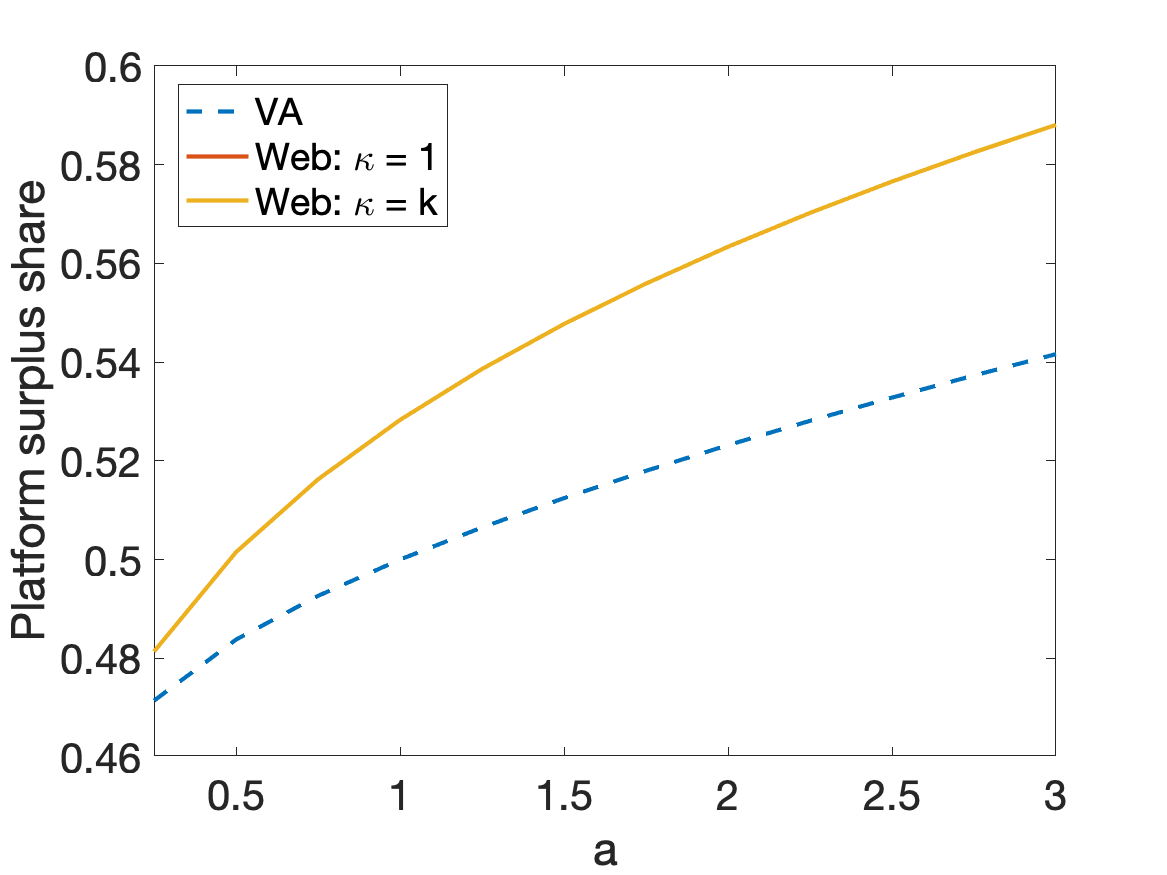}}
    \subfloat[$\rho=1, k=6.$ ]{\includegraphics[width=0.32\textwidth]{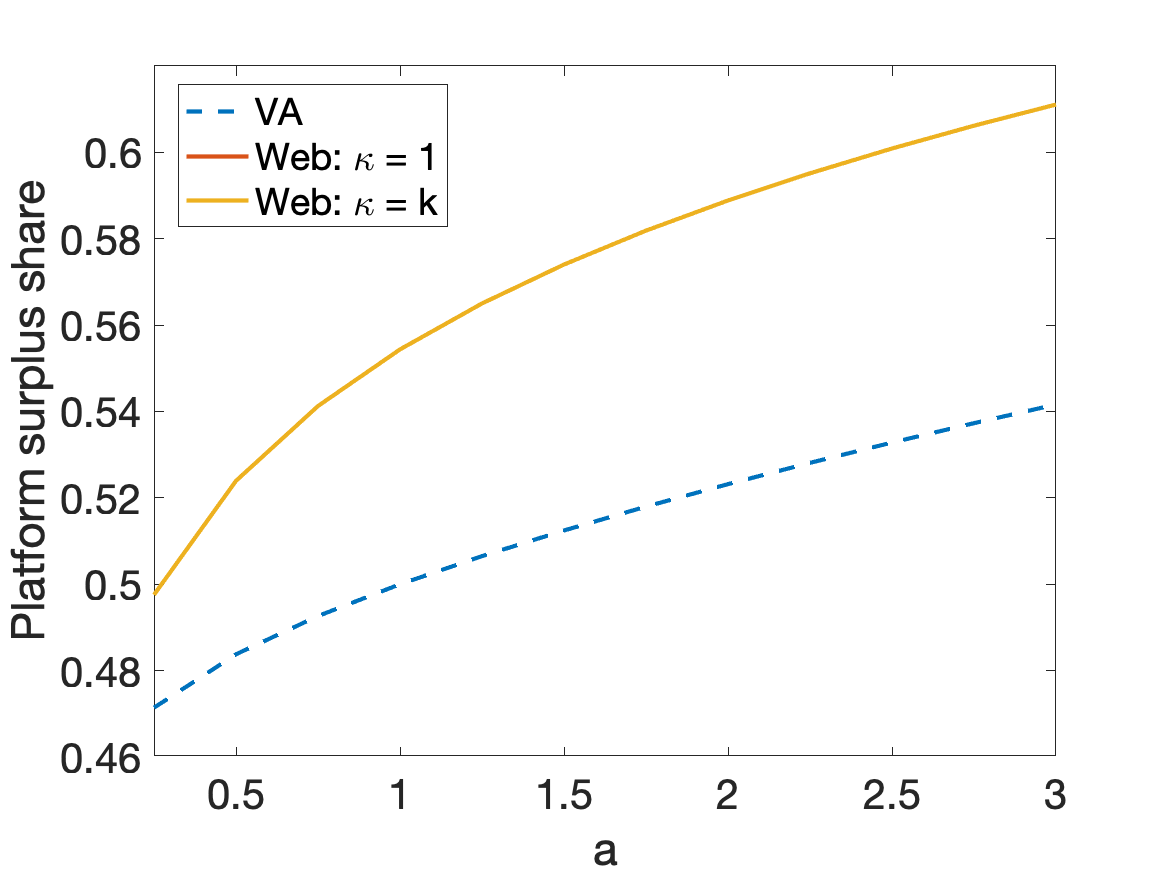}}
    \caption{seller surplus share as a function of the Gamma shape parameter $a$ for fixed mean $ab = 1$ and $N=6$ products for different values of $\rho$ and $k$ and for $\kappa = 1$ or $k$. As $a$ increases, the variance of the private valuations decreases.}
    \label{webvsva_surplusallocation}
    \end{figure}

   To compare the surplus allocation under the two regimes, we consider again the Gamma-distributed private valuations studied in Section \ref{sec:implication} with 
        %  \noindent \textbf{Exponential distribution.} \qquad In figure \ref{web_surplussplit}, we observe that as number of product per page increases, seller gets more surplus share. This is intuitively clear, as when product per page $k$ increases, web-interface has more power of commitment on pricing, and it grabs more surplus split compared with virtual assistant seller. In virtual assistant seller ($k=1$), surplus split is $50-50$. 
    % \noindent \textbf{Gamma distribution.}
   shape parameter $a$, scale parameter $b$, and constant mean $ab=1$. We increase the shape parameter $a$ from $ 0.5$ to $3$, decreasing the variance $\frac{1}{a}$ and thereby the amount of private information. Figure \ref{webvsva_surplusallocation} shows the seller's surplus share (the ratio $\frac{V_p}{V_p+V_c}$ of seller profit to total surplus) for $N=6$ products for different values of the impatience parameter $\rho$, number of products on a page $k$ and acceleration factor $\kappa$.  
    
    We first consider how the problem parameters (the Gamma shape parameter $a$, which determines the amount of private information; the acceleration factor $\kappa$, which affects the consumer's speed of evaluation; and the number of products presented per page $k$) affect the surplus allocation under the web-interface. 
    
    Figures \ref{webvsva_surplusallocation} show that the seller surplus share is decreasing in the shape parameter $a$. This is consistent with our prior intuition and results for the virtual assistant case (Figure \ref{fig:va_info_surplusallocation} in Section \ref{sec:implication}): as $a$ increases, the amount of private information decreases, which allows the seller to extract more of the total surplus.
    
    Second, as the acceleration factor $\kappa$ increases, the consumer can explore more products, which benefits both the consumer and the seller. As a result, the \emph{levels} of both the consumer surplus and the seller profit increase. Figures \ref{webvsva_surplusallocation} show that this acceleration has only a small effect on the surplus \emph{ratio}. 
    
    Finally, looking across Figures \ref{webvsva_surplusallocation} (with the same $a$), we observe that the seller's surplus share increases with $k$. This effect is best interpreted in conjunction with a comparison of the web-interface and the virtual assistant.  
    A key driver of this comparison is the seller's power of price commitment under the web-interface. A well-known fundamental result in game theory is that being able to commit to a strategy before other players move is generally beneficial (cf. \cite{courty2000sequential}). 
    % **** THE FOLLOWING SENTENCE MAKES NO SENSE TO ME; WE'LL NEED TO REDO THE DISCUSSION HERE **** Without such a commitment, the buyers may anticipate that the object will be offered for sale again in a later period with lower price. Hence, it gives the buyers more bargaining power (cf. \cite{krishna2009auction}).
    
    When comparing the web-interface to the virtual assistant, the former enables the seller to commit to a full page of prices while the virtual assistant dynamically prices a product at a time. Thus, the web interface has higher commitment power, which is an increasing function of the number of products per page $k$. Thus, we expect \emph{(i)} the web-interface to result in a higher surplus share for the seller, and \emph{(ii)} the surplus share to increase with $k$, which is what we observe in Figure \ref{webvsva_surplusallocation}.  As for \emph{(ii)}, an increase in $k$ also allows the consumer to view more products, but as we have seen earlier, the effect of this factor on the surplus \emph{share} is small.    
        
    % When the player can make a-prior In the seqeuential case, it has the benefit of ... but actually it loses ...
    
    % 1. trade-offs, web > va 
    However, the viewing rate is always higher in a web interface (web interface: $\tau_p\kappa$, virtual assistant $\tau_p$), which enables the consumer to view more products in a web interface. As we discussed before, it favors both consumer and the seller, thus has little influence on the \emph{surplus allocation}. Overall, the effect of commitment power dominates the effect of  viewing rate and as we observe from figure \ref{webvsva_surplusallocation}, the surplus allocation is always higher in a web-interface. 

%% file: sec7_conclusion.tex
\section{Concluding remarks}\label{sec:conclusion}
With an increasing number of consumers choosing to purchase products through virtual assistants, this emerging channel is expected to become an important gateway to commerce.  It is important to understand how the special features of virtual assistants (in particular, the sequential nature of product presentation) affect the market outcomes.  

In this paper, we developed a model of a forward-looking consumer who strategically makes sequential purchase decisions after submitting a request to a virtual assistant which makes ranking and pricing decisions. The virtual assistant may operate on behalf of a profit-maximizing seller, it may be altruistic, or it may act as a consumer agent. We find the optimal prices under a general private valuation distribution and derive them in closed-form when the distribution is exponential. We find that in the exponential case, a profit-maximizing seller extracts the same surplus as the consumer. As a result, the profit-maximizing ranking also maximizes the consumer surplus. For an altruistic seller or a consumer agent, we find that pricing at cost is optimal and the consumer extracts the entire surplus. We develop algorithms for optimally ranking products and find that the simple descending or ascending rankings are optimal when consumers are highly patient or impatient. We propose the double-rank approximation algorithm which is shown to capture at least $95\%$ of the surplus in numerical experiments.

In July 2020, the European Commission launched an inquiry into the market for consumer products and services linked to the Internet of Things with a focus on voice assistants. Citing the ``incredible potential'' of these devices, Commission Executive Vice-President Margrethe Vestager focused on the ``risk that some of these players could become gatekeepers of the Internet of Things, with the power to make or break other companies. And these gatekeepers might use that power to harm competition, to the detriment of consumers... whether that's for a new set of batteries for your remote control or for your evening takeaway. In either case, the result can be less choice for users, less opportunity for others to compete, and less innovation'' (\cite{Vestager20}).

Analyzing the effects of selling through a VA compared to selling through a traditional web interface calls for an explicit analysis of the two equilibria. We perform such an analysis where the seller may sell through a VA or through a web interface, where products are presented on multiple web pages with each page showing multiple products simultaneously. One might expect that when the VA maximizes the seller's profits, it will exploit its gatekeeper control over product presentation to extract a larger share of the surplus. We find that the opposite  is true: the seller's equilibrium surplus share is in fact larger with a web interface, where it can credibly commit to fixed prices on each page. We also find that when the consumer's private valuations are exponentially distributed, the optimal prices within a page are decreasing in product valuations. 

There are several interesting extensions to this work. First, it will be useful to extend our analysis to the case where the consumer has imperfect recall. Second, it is interesting to study a three-sided platform model where strategic suppliers and consumers are mediated through a VA. The VA might first decide on the price of each position. Observing these prices, suppliers might then decide which positions to acquire and bring in their products with associated prices. Finally, consumers might use the platform to make buying decisions.

%% file: app1_notation.tex
\section{Notation}\label{app_notation}
\begin{itemize}
    \item $i$ : Product index; $i=1,2,\dots,N$, where $N$ is the number of products.
    \item $\tau_c$ :  a consumer request is live for an exponential duration with rate $\tau_c$.
    \item $\tau_p$ : the presentation and evaluation time of each offer is exponentially distributed with rate $\tau_p$.
    \item $v_i+\epsilon_i$ : valuation of product $i$, where
    \item $v_i$ : observable part of the valuation of product $i$;
    \item $\epsilon_i$ : unobservable (random) part of the valuation of product $i$,  $\epsilon_i\sim F$.
    \item $\alpha$ :  when $\epsilon_i$ are exponentially distributed, $\EE(\epsilon_i)=1/\alpha$.
    \item $\bsigma$  : product ranking permutation, $\bsigma=(\sigma_1,\dots,\sigma_n)$. 
    \item $\rho$ : probability that the consumer completes a product evaluation.
    \item $p_i $ : consumer price of product $ i$.
    \item $c_i$  : platform cost to acquire and ship product $i$ when it's sold to the consumer.
    \item $\delta_i$ : consumer threshold valuation. The consumer buys product $i$ when $u_i=v_i-p_i+\epsilon_i\geq\delta_i$.
    \item $\PP_i=\PP(\epsilon_{\sigma_i} \geq p_{\sigma_i}+\delta_{\sigma_i}-v_{\sigma_i})$ : probability that the consumer buys the $i^{th}$ product offered.  
\end{itemize}

\section{Parameters in Numerical Examples}\label{app_numerical_setting}

%For details of numerical example see appendix B.1
The numerical examples in sections \ref{sec:optimal_ranking}--\ref{sec:web_vs_va} consider the following default setting:
\begin{itemize}
    \item $N = 6$ products with valuation vector $v = (0, \frac{1}{6}, \frac{1}{3}. \frac{1}{2}, \frac{2}{3}, \frac{5}{6})$ and costs $c_i = 0  (i=1,2,\dots,6)$; and 
    \item $\epsilon \sim $ Gamma $(a, b)$.
\end{itemize}

Gamma $(a, b)$ is the Gamma distribution with shape parameter $a$ and scale parameter $b$. This distribution is widely used in applications; we summarize here some of its salient properties: 
\begin{itemize}
    \item Mean $\EE(\epsilon)=ab$;
    \item Variance $Var(\epsilon)=ab^2$;
    \item For a fixed mean, increasing the shape parameter $a$ reduces the variance $ab^2=\EE(\epsilon)/a$.
    \item $\epsilon$ is exponential when $a=1$ and has the uniform distribution as its limit as $a\to \infty$. \\
  \end{itemize}

%% file: app2_proofs.tex
\section{Proofs}

% \subsection*{Section \ref{sec:equlibrium_pricing}} 

\subsubsection*{Proof of Proposition \ref{prop:pp_eqm_general}}

The proof is by backward induction.

\begin{enumerate}[label=(\alph*),font=\itshape]
\item Stage $N$: In stage $N$, the consumer will purchase if and only if 
\[ \epsilon_N + v_N - p_N \geq \delta_N=0,  \]
and the probability of purchase at any price $p$ is $1-F(p-v_N)$. The seller's expected profit is then given by $(1-F(p-v_N)) \cdot (p-c_N)$; denote the maximizer by $p^*_N$. 

The maximizer exists because both at $p_N=c_N$ and as $p_N\to\infty$, the expected profit goes to zero.
% so we are maximizing a continuous function over a bounded support. 
The optimal expected profit is positive since $p_N=c_N$ results in zero profit, and the derivative at $c_N$ (given by $(1-F(c_N-v_N))$ is positive (since $(v_i - c_i )$ are non-positive and $F(\cdot)$ is positively supported).
\\
The equilibrium probability of purchase in stage $N$ is clearly \[\PP_N=1-F(p^*_N-v_N),\] and the seller's expected profit is thus
\eqss{
V^p_N =\PP_N (p^*_N-c_N).
}

The consumer's expected surplus under the optimal price is equal to the consumer's expected valuation times the purchase probability conditional on the valuation exceeding the stage-$N$ threshold, hence  
\eqss{
V_N^c =  \PP_N \EE(\epsilon_N + v_N-p^*_N|\epsilon_N +v_N-p^*_N\geq \delta_N).
}

\item We next use backward induction. In stage $i=N-1,\dots,1$, the consumer will purchase if and only if her current-stage realized valuation is greater than her expected future return, i.e., if $\epsilon_{i} + v_{i} - p_{i} \geq  \rho V_{i+1}^c$, which gives us a threshold-policy structure with $\delta_i=\rho V_{i+1}^c$.

It follows that the probability of purchase given any price $p$ is $1-F(p-v_i+\rho V^c_{i+1})$. The seller's expected profit is the sum of its expected gain when the consumer accepts the current offer and its expected gain when the consumer continues to the next stage, given by:
    \[ (1-F(p-v_{i}+\rho V_{i+1}^c)) (p-c_{i})+  F(p-v_{i}+\rho V_{i+1}^c) \rho  V_{i+1}^p.\]
A similar argument to the one for $i=N$ shows that a finite maximizer always exists, the equilibrium probability of purchase in stage $i$ is $\PP_i=1-F(p^*_i-v_i+\rho V^c_{i+1})$, the seller's expected profit in stage $i$ is 
    \[
 V_{i}^p =  \PP_{i} (p^*_{i}-c_{i}) + (1-\PP_{i}) \rho V_{i}^p ,
\]
 and the expected consumer surplus is 
\[ V_{i}^c=  \PP_{i} \EE(\epsilon_{i} + v_{i}-p^*_{i}|\epsilon_{i}+v_i-p^*_i \geq \delta_i) ) + (1-\PP_{i}) \rho V_{i+1}^c  .  \]
\end{enumerate}

\subsubsection*{Proof of Corollary \ref{cor:prop_pp_eqm_exp}}\label{pf_prop_pp_eqm_exp}

The results follow from Proposition \ref{prop:pp_eqm_general} with $\epsilon\sim\exp(\alpha)$.

\begin{enumerate}[label=(\alph*),font=\itshape]
\item  In stage $N$, $\delta_N = 0$ and since $\epsilon\sim\exp(\alpha)$, the seller's expected profit is given by $(1-F(p-v_N)) (p-c_N)$ with optimal price \eqss{
p^*_N= c_N+\dfrac{1-F(p_N-v_N)}{f(p_N-v_N)}=c_N+\dfrac{1}{\alpha}.
}
Plugging $p^*_N$ into the expression for $\PP_N$, we have 
\eqss{
\PP_N=\exp(-\alpha(c_N+\dfrac{1}{\alpha}-v_N)) = \exp(\alpha(v_N-c_N)-1)
}
and 
\eqss{V^p_N =\PP_N (p^*_N-c_N)=\dfrac{1}{\alpha}\exp(\alpha(v_N-c_N)-1)=\dfrac{1}{\alpha}\PP_N.}
Now, plugging $p^*_N$ and $\delta_N$ into the expression for $V^c_N$ and using the memeoryless property of the exponential distribution, we get 
\eqss{
V_N^c &= \PP_N  \EE(\epsilon_N + v_N-p^*_N|\epsilon_N +v_N-p^*_N\geq \delta_N)\\
&=\dfrac{1}{\alpha}\exp(\alpha(v_N-c_N)-1)=\dfrac{1}{\alpha}\PP_N.
}

\item For stages $i=N-1,\dots,1$, Proposition \ref{prop:pp_eqm_general} implies $\delta_i=\rho V_{i+1}^c$ with optimal price 
    \eqss{p^*_i&=\arg\max_{p} \ (1-F(p-v_{i}+\rho V_{i+1}^c)) (p-c_{i})+  F(p-v_{i}+\rho V_{i+1}^c) \rho  V_{i+1}^p\\
     &=c_{i}+\dfrac{1-F(\delta_i)}{f(\delta_i)}+\rho V{p}_{i+1}=c_i+\dfrac{1}{\alpha}+\rho V^{p}_{i+1}.}
     The equilibrium probability of purchase in stage $i$ is given by
     \eqss{\PP_i=1-F(p^*_i-v_i+\rho V^c{i+1})=\exp(\alpha(v_i-c_i)-1)\exp(-\alpha\rho(V^c_{i+1}+V^{p}_{i+1})),}
the expected seller profit is 
  \eqss{
 V_{i}^p &=  \PP_{i} (p^*_{i}-c_{i}) + (1-\PP_{i}) \rho V_{i+1}^p \\
 &=\dfrac{1}{\alpha}\PP_i+\rho V^{p}_{i+1}=\dfrac{1}{\alpha}\sum^{N}_{k=i}\rho^{k-i}\PP_k
,}
and the expected consumer surplus is  
\eqss{ V_{i}^c&=  \PP_{i} \EE(\epsilon_{i} + v_{i}-p^*_{i}|\epsilon_{i}+v_i-p^*_i \geq \delta_i) ) + (1-\PP_{i}) \rho V_{i+1}^c  \\
& = \dfrac{1}{\alpha}\PP_i+\rho V^{c}_{i+1}=\dfrac{1}{\alpha}\sum^{N}_{k=i}\rho^{k-i}\PP_k.}
\end{enumerate}

\subsubsection*{Proof of Proposition \ref{prop:apca_eqm}}\label{pf_prop_apca_eqm}

We prove proposition \ref{prop:apca_eqm} by backward induction. 
\begin{enumerate}[label=(\alph*),font=\itshape]
\item In stage $N$, the optimal threshold is $\delta_N = 0$ and the total surplus is 
\eqss{
V_N^s & = \EE(\epsilon_N + v_N - c_N | \epsilon_N + v_N - p_N) \PP( \epsilon_N + v_N - p_N \geq \delta_N) \\
& =  \int_{\delta_N + p_N- v_N}^{\infty} (x + v_N - c_N) f(x) \dd x .}
Now, \eqss{ \pap{V_N^s}{p_N} & = - (\delta_N + p_N -c_N) f(\delta_N + p_N -v_N) = - (  p_N -c_N) f(p_N -v_N). 
}
Since $F(\cdot)$ has positive support and $ v_N \leq c_N$, $V_N^s$ is constant in $(-\infty, v_N)$, strictly increasing in $ (v_N, c_N]$, and strictly decreasing in $ (c_N, \infty).$ Thus, the optimal price is $ p_N = c_N,$ $V_N^s = V_N^c$ and $V_N^p = 0.$\\

\item In stage $i=N-1,\dots,1$, $\delta_i = \rho V_{i+1}^c =\rho V_{i+1}^s$, the expected total surplus is 
\eqss{
V_i^s & = \EE(\epsilon_i + v_i - c_i| \epsilon_i + v_i - p_i \geq \delta_i) \PP( \epsilon_i + v_i - p_i \geq \delta_i) + (1- \PP( \epsilon_i + v_i - p_i \geq \delta_i)) \rho V_{i+1}^s \\
& =  \int_{\delta_i + p_i- v_i}^{\infty} (x + v_i - c_i) f(x) \dd x + (1- \int_{\delta_i + p_i- v_i}^{\infty}f(x) )  \rho V_{i+1}^s\\
&= \int_{\delta_i + p_i- v_i}^{\infty} (x + v_i - c_i - \rho V_{i+1}^s  ) f(x) \dd x + \rho V_{i+1}^s ,}

and \eqss{
\pap{V_i^s}{p_i} & = - (\delta_i - \rho V_{i+1}^s  p_i -c_i) f(\delta_i + p_i -v_i) = - ( p_i -c_i) f(\delta_i + p_i -v_i) 
.}

Again, $V_i^s$ is constant in $(-\infty, v_i - \delta_i)$, strictly increasing in $ ( v_i - \delta_i, c_i]$, and strictly decreasing in $ (c_i, \infty).$ Thus, the optimal price is $ p_i = c_i$ and the inductive hypothesis holds. 

Finally, since $V^i_p=0$ for all $i$, $p_i=c_i$ maximizes both the expected consumer surplus and the total surplus. 

\end{enumerate}

% 3. limiting distribution 

% \subsection*{Section \ref{sec:optimal_ranking}}  
\subsubsection*{Proof of Proposition \ref{prop_ExtRho}}\label{pf_prop_ExtRho}
%\commwb{Make it consistant with the statement}
 (a) We show that there exists a $\rho_0>0$ such that for all $\rho\in[0,\rho_0)$, the descending ranking (in value margins) is optimal. We first prove the result for a profit-maximizing seller and then for an altruistic seller or a consumer agent. \\
 
 Let the ranking permutation be $\sigma$. For a profit-maximizing VA, following Corollary \ref{cor:prop_pp_eqm_exp}, the seller's expected profit in the first stage is given by
 $$V^p_1=\dfrac{1}{\alpha}\sum^{N}_{k=1}\rho^{k-1}\PP_{k}.$$
 Taking $\rho\to0$, as $\PP_k$ are bounded for all $k$, the the expected first-stage profit is given by 
 $$V^p_1=\frac{1}{\alpha}\PP_1+o(\rho)$$
 % And the product presented first drives the entire profit. 
 by Corollary \ref{cor:prop_pp_eqm_exp}. Let $ M_{\sigma(i)} = \exp(\alpha(v_{\sigma(i)}-c_{\sigma(i)})$. Then, 
 $$\PP_1(\rho)=\exp(\alpha (v_{\sigma(1)}-c_{\sigma(1)})-1)\exp(-\alpha\rho (2V^{p}_2))=M_{\sigma(i)}\exp(-1)\exp(-\alpha\rho (2V^{p}_2)).$$
 Taking $\rho\to0$ and using the Taylor's series expansion for $\exp(-\alpha\rho (2V^{p}_2))$, we have
  $$\PP_1(\rho)=M_{\sigma(i)}\exp(-1)-2\alpha V^p_2\rho+o(\rho^2).$$
 Since $V^p_2$ is bounded, 
 % $\exp(\alpha (v_{\sigma(1)}-c_{\sigma(1)})-1)$ drives the entire profit at the first stage, and
  $$V^p_1=M_{\sigma(i)}\exp(-1)+o(\rho).$$
Since $M_{\sigma(i)}$ is increasing in $v_{\sigma(i)}-c_{\sigma(i)}$, the VA maximizes $V^p_1$ by placing the product with the highest value-margin first.
\\
It is of course possible that the consumer will progress to the second stage (due to a small enough realization of $\epsilon_1$). By Corollary \ref{cor:prop_pp_eqm_exp}, the corresponding expected profit is 
$$V^p_2=\dfrac{1}{\alpha}\sum^{N}_{k=2}\rho^{k-2}\PP_k.$$
Similar to our analysis of stage $1$, 
$$V^p_2 = M_{\sigma(2)}\exp(-1)+o(\rho),$$
and the optimal $\sigma(2)$ is given by the remaining product with the highest value-margin, namely the one with the second-highest value margin. By induction, the optimal ranking as $\rho\to0$ is thus descending in products' value margins. 
By Corollary \ref{cor:prop_pp_eqm_exp}, the optimal prices are given by
 $$p^*_i=c_i+\frac{1}{\alpha}+\rho V^p_{i+1} = c_i+\frac{1}{\alpha}+o(\rho),$$
since $V^p_{i+1}$ are bounded.\\
\\

We next consider the case of a consumer agent VA. Let the ranking permutation be $\sigma$. By Corollary \ref{cor:apca_pricing}, the expected first-stage surplus is given by
$$V^c_{1} =  \PP_{1} \EE(\epsilon + v_{\sigma(1)}-c_{\sigma(1)}|\epsilon + v_{\sigma(1)}-c_{\sigma(1)}\geq \rho V^c_2 ) + (1-\PP_{1})  \rho V_{2}^c.$$
As $(1-\PP_1)V^c_2$ is bounded from above, taking $\rho\to0$, 
$$V^c_{1} =  \PP_{1} \EE(\epsilon + v_{\sigma(1)}-c_{\sigma(1)}|\epsilon + v_{\sigma(1)}-c_{\sigma(1)}\geq \rho V^c_2 ) + o(\rho)$$
$$=(1-F(-(v_{\sigma(1)}-c_{\sigma(1)})))\EE(\epsilon+v_{\sigma(1)}-c_{\sigma(1)}|\epsilon+v_{\sigma(1)}-c_{\sigma(1)}\geq 0)+o(\rho).$$

For $t=v_i-c_i$, 
$$\dfrac{\partial }{\partial t}(1-F( - t))\int^{\infty}_{-t}(\epsilon+t)f(\epsilon)d\epsilon$$
$$=(1-F(-t))\int^{\infty}_{-t}f(\epsilon)d\epsilon+f(-t) \int^{\infty}_{-t}(\epsilon+t)f(\epsilon)d\epsilon \geq0,$$
so $V^c_1$ is increasing in the value margin of the first product presented. To maximize $V^c_1$, the VA should thus place the item with the highest value-margin first.
\\
% It is possible that the consumer will progress to the second stage (due to a small enough realization of $\epsilon_1$). 
Proceeding to stage $2$, from Corollary \ref{cor:apca_pricing} the expected consumer surplus is given by 
$$V^c_{2} =  \PP_{2} \EE(\epsilon + v_{\sigma(2)}-c_{\sigma(2)}|\epsilon + v_{\sigma(2)}-c_{\sigma(2)}\geq \rho V^c_3 ) + (1-\PP_{2})  \rho V_{3}^c.$$
As before, as $\rho\to0$ $V^c_2$ is dominated by the expected consumer surplus at stage 2:
$$V^c_2=(1-F(-(v_{\sigma(2)}-c_{\sigma(2)})))\EE(\epsilon+v_{\sigma(2)}-c_{\sigma(2)}|\epsilon+v_{\sigma(2)}-c_{\sigma(2)}\geq 0)+o(\rho).$$
Following the foregoing analysis, since $V^c_2$ is increasing in the value-margin of the second product, the $\sigma(2)$ that maximizes $V^c_2$ is given by the product with second-highest value margin. By induction, the optimal ranking as $\rho\to0$ is given by descending order of the value-margins.\\
\\
By Corollary \ref{cor:apca_pricing}, the same results hold for an altruistic VA. 
\\
\\
(b) We prove first the case of profit-maximization, using the following claim. \\
\\
\textbf{Claim:} If $\sigma =  (\sigma(1),\cdots, \sigma(N)) $ is the optimal ranking, the sub-permutation $\Bar{\sigma }^i:= (\sigma(i),\sigma(i+1), \cdots, \sigma(N))$ is the optimal ranking for the subproblem from stage $i$ through $N$, i.e., $ V_i^p(\Bar{\sigma }^i) = \max_{\sigma'} V_i^p(\sigma') $.\\
\\
\begin{pf} Let $M_i = \exp(\alpha (v_i - c_i))$ and rewrite $V_i^p$ as a function of $M_{\sigma(i)}$ and $V_{i+1}^p: V_i^p = f(M_{\sigma(i)}, V^p_{i+1})$. Now, 
\eqs{ \label{eq: ViMono} 
V_i^p = f(M_{\sigma(i)}, V^p_{i+1}) & = \PP_i (p_{\sigma(i)} - c_{\sigma(i)}) + (1-\PP_i ) \rho V^p_{i+1} \\ \nonumber
 & = \PP_i (\frac{1}{\alpha} + \rho V^p_{i+1}  ) + (1-\PP_i ) \rho V^p_{i+1}\\ \nonumber
 & = \frac{1}{\alpha} \exp(\alpha(v_{\sigma(i)} -c_{\sigma(i)}) - 1 ) \exp(-2\alpha\rho V^p_{i+1}) +  \rho V^p_{i+1}\\ \nonumber
 & =  \frac{1}{\alpha e} \exp(-2\alpha\rho V^p_{i+1}) M_{\sigma(i)}  + \rho V^p_{i+1}.}
We have that $ \pap{f}{V} = \rho (1-\frac{2}{e}e^{-2\alpha V} M)> 0$, which implies that $V^p_i$ is increasing in $ V^p_{i+1}$. \\

Now, when we fix the order of the first $i-1$ products, $ \pap{V^p}{V_i^p} = \pap{V^p}{V^p_2} \pap{V^p_2}{V^p_3} \cdots \pap{V^p_{i-1}}{V^p_i} > 0 $. Therefore, if $\Bar{\sigma }^i$ does not maximize $V_i^p$, there exists another $\Bar{\sigma'}^i$ such that  $V_i^p(\Bar{\sigma'}^i)> V_i^p(\Bar{\sigma }^i) $. But then
\[ V^p( (\sigma(1), \cdots,  \sigma(i-1), \Bar{\sigma' }^i ) )> V^p( (\sigma(1), \cdots,  \sigma(i-1), \Bar{\sigma }^i) )  =V^p(\sigma ),  \]
thereby contradicting the optimality of $\sigma.$ 
\end{pf}
\\
\\
We next show by backward induction that for $\rho = 1$ and an optimal ranking $ \sigma$, each sub-permutation $\Bar{\sigma }^i$ satisfies the monotonicity condition $v_{\sigma(i)} - c_{\sigma(i)} \leq v_{\sigma(i+1)} - c_{\sigma(i+1)} \leq \cdots \leq v_{\sigma(N)} - c_{\sigma(N)}$, or equivalently
% $ With $M_i = \exp(\alpha (v_i - c_i))$, we now show that 
$M_{\sigma(i)} \leq M_{\sigma(i+1)} \leq \cdots \leq M_{\sigma(N)} $ for all $i$.
\begin{itemize}
% \item \textbf{Inductive hypothesis}: If the optimal ranking is $\sigma $, for any $i$, the sub-permutation $ \Bar{\sigma }^i$ satisfies that $M_{\sigma(i)} \leq M_{\sigma(i+1)} \leq \cdots \leq M_{\sigma(N)}.$
\item \textbf{Baseline case:} The statement is trivially true for the baseline case when $i=N$.

\item \textbf{Induction:} Assume the statement holds for for $k=i+1, \cdots, N$. We prove it also holds for $k=i,$ namely $M_{\sigma(i)} \leq M_{\sigma(i+1)}$.

Assume by contradiction that $M_{\sigma(i)} > M_{\sigma(i+1)}$. Then, we switch products $\sigma(i)$ and $\sigma(i+1)$ to obtain the permutation $\sigma':$
\[\sigma' = (\sigma(1), \sigma(2), \cdots, \sigma(i+1), \sigma(i), \sigma(i+2), \cdots, \sigma(N)).\]
We will show that the sub-permutation $\Bar{\sigma }^i:= \{\sigma(i), \sigma(i+1), \cdots, \sigma(N)\} $ doesn't maximize $V^p_i$, contradicting the inductive hypothesis. \\
\\
Since $\sigma $ and $\sigma'$ only differ in the $i-$th and $(i+1)$-th positions, $\Bar{\sigma}^{i+2} =\Bar{\sigma'}^{i+2}$ and we denote $V_{i+2}^p(\Bar{\sigma}^{i+2}) = V_{i+2}^p(\Bar{\sigma'}^{i+2}) $ by $\Bar{V}.$ The expected profits under the two rankings are as follows:    
\eqss{
    V^p_{i+1}(\Bar{\sigma}^{i+1})  & = \Bar{V} +\frac{M_{\sigma(i+1)}}{\alpha e} e^{-2\alpha \Bar{V}} ,\\
    V^p_{i}(\Bar{\sigma}^i) & =  \Bar{V} +\frac{ M_{\sigma(i+1)}}{\alpha e} e^{-2 \alpha \Bar{V}} + \frac{M_{\sigma(i)}}{\alpha e} e^{-2\alpha  \rho \Bar{V} -\frac{2}{e} e^{-2\alpha \Bar{V}} M_{\sigma(i+1)}} ,\\
    V^p_{i+1}(\Bar{\sigma'}^{i+1})  & =\Bar{V} +\frac{M_{\sigma(i)}}{\alpha e} e^{-2\alpha  \Bar{V}} ,\text{~and}\\
    V^p_{i}(\Bar{\sigma'}^{i}) & =  \Bar{V} +\frac{ M_{\sigma(i)}}{\alpha e} e^{-2 \alpha \Bar{V}} + \frac{M_{\sigma(i+1)}}{\alpha e} e^{-2\alpha   \Bar{V} -\frac{2}{e} e^{-2\alpha \Bar{V}} M_{\sigma(i)}} .
    }
    
    % \eqss{
    % V^p_{i+1}(\Bar{\sigma}^{i+1})  & = \rho \Bar{V} +\frac{M_{\sigma(i+1)}}{\alpha e} e^{-2\alpha \rho \Bar{V}} ,\\
    % V^p_{i}(\Bar{\sigma}^i) & = \rho^2 \Bar{V} +\frac{\rho M_{\sigma(i+1)}}{\alpha e} e^{-2\rho^2 \alpha \Bar{V}} + \frac{M_{\sigma(i)}}{\alpha e} e^{-2\alpha  \rho \Bar{V} -\frac{2\rho}{e} e^{-2\alpha \Bar{V}} M_{\sigma(i+1)}} ,\\
    % V^p_{i+1}(\Bar{\sigma}^{i+1}')  & = \rho \Bar{V} +\frac{M_{\sigma(i)}}{\alpha e} e^{-2\alpha \rho \Bar{V}} , \text{and}\\
    % V^p_{i}(\Bar{\sigma}^{i}') & = \rho^2 \Bar{V} +\frac{\rho M_{\sigma(i)}}{\alpha e} e^{-2\rho^2 \alpha \Bar{V}} + \frac{M_{\sigma(i+1)}}{\alpha e} e^{-2\alpha  \rho \Bar{V} -\frac{2\rho}{e} e^{-2\alpha \Bar{V}} M_{\sigma(i)}} .
    % }
   Showing that $V^p_{i}(\Bar{\sigma}^i) < V^p_{i}(\Bar{\sigma'}^i)$ is equivalent to showing that the function
   \[g(x): =  \frac{1}{x} - \frac{\exp( - \frac{2}{e}\exp(-2\alpha \Bar{V})x)}{x} \]
   is increasing in $x$. Let $\frac{2}{e}\exp(-2\alpha \Bar{V}) =: k<1$, then
   \[ g(x)=  - \frac{1}{x} + \frac{\exp(- k x)}{x}   . \]
   The first derivative of $g(x)$ is  
   \eqss{
   g'(x)   & = \frac{1-\exp(-kx)(kx+1)}{x^2}.
   %\frac{1}{x^2} - \frac{ kx\exp(-kx) + \exp(-kx)       }{x^2}\\
   }
   Since $ \exp(-kx)(kx+1) < 1$ for $0< kx <1$, it follows that $g'(x) >0$ and $V^p_{i}(\Bar{\sigma}^i) < V^p_{i}(\Bar{\sigma'}^i)$. This contradicts the claim that $\Bar{\sigma}^i$ maximizes $V^p_{i}$.  \\
   \\
From equation \eqref{eq: ViMono} we know that  $V^p_i(\sigma) and V^p_i(\sigma')$ are continuous in $\rho.$ By continuity, there exists a $\rho_1$ such that the monotonicity would hold for all $\rho \in (\rho_1, 1]$. We have thus proved that there exists a $\rho_1$ such that for all $\rho \in (\rho_1, 1]$, $ M_{\sigma(i)} < M_{\sigma(i+1)} <\cdots < M_{\sigma(N)}$.
\end{itemize}
\noindent
We next prove the result for an altruistic VA and a consumer agent. Notice that in the proof of the profit-maximizing case, the key was to show that \textit{(i)} $V^p_i$ is monotone in $V^p_{i+1}$, and \textit{(ii)} $V^p_i(\Bar{\sigma}^i)< V^p_i(\Bar{\sigma'}^i)$ when $\rho=1$. To prove \textit{(i)}, we have for a consumer agent (and an altruistic platform) with $\rho=1$:
\eqss{ V^p_i = f(M_{\sigma(i)}, V_{i+1}) &=  \frac{M_{\sigma(i)}}{\alpha} \exp(-\alpha \rho V^p_{i+1}) +  \rho V^p_{i+1}, and\\
\pap{f}{V^p_{i+1} } & =  \rho - \rho M_{\sigma(i)} \exp(-\alpha \rho V^p_{i+1}) >0}
which proves \textit{(i)}.
Similarly,
\eqss{
V^p_{i}(\Bar{\sigma}^i) &  =  \Bar{V} + \frac{ M_{\sigma(i)}}{\alpha}\exp(-\alpha  \Bar{V}) + \frac{M_{\sigma(i)}}{\alpha} \exp( - \alpha \Bar{V} -  M_{\sigma(i+1)} \exp(-\alpha  \Bar{V}  )),\\
V^p_{i}(\Bar{\sigma'}^i) &  = \Bar{V} + \frac{ M_{\sigma(i)}}{\alpha}\exp(-\alpha  \Bar{V}) + \frac{M_{\sigma(i+1)}}{\alpha} \exp( - \alpha \Bar{V} -  M_{\sigma(i)} \exp(-\alpha   \Bar{V}  )), and \\
\frac{\alpha}{M_{\sigma(i)}M_{\sigma(i+1)} } ( V_{i}(\Bar{\sigma'}^i) -V_{i}(\Bar{\sigma}^i))&= \frac{1}{M_{\sigma(i)}}\exp( - \alpha \Bar{V} - M_{\sigma(i)} \exp(-\alpha  \Bar{V}  )) -   \frac{1  }{M_{\sigma(i)}}\exp(-\alpha \Bar{V}) \\
& - \frac{1}{M_{\sigma(i+1)}}\exp( - \alpha  \Bar{V} -  M_{\sigma(i+1)} \exp(-\alpha   \Bar{V}  )) -   \frac{1  }{M_{\sigma(i+1)}}\exp(-\alpha  \Bar{V}.) 
'}

% \eqss{
% V^p_{i}(\Bar{\sigma}^i) &  = \rho^2 \Bar{V} + \frac{\rho  M_{\sigma(i)}}{\alpha}\exp(-\alpha \rho \Bar{V}) + \frac{M_{\sigma(i)}}{\alpha} \exp( - \alpha \rho^2 \Bar{V} - \rho M_{\sigma(i+1)} \exp(-\alpha \rho  \Bar{V}  ))\\
% V^p_{i}(\Bar{\sigma}^i') &  = \rho^2 \Bar{V} + \frac{\rho  M_{\sigma(i)}}{\alpha}\exp(-\alpha \rho \Bar{V}) + \frac{M_{\sigma(i+1)}}{\alpha} \exp( - \alpha \rho^2 \Bar{V} - \rho M_{\sigma(i)} \exp(-\alpha \rho  \Bar{V}  )) \\
% \frac{\alpha}{M_{\sigma(i)}M_{\sigma(i+1)} } ( V_{i}(\Bar{\sigma}^i') -V_{i}(\Bar{\sigma}^i))&= \frac{1}{M_{\sigma(i)}}\exp( - \alpha \rho^2 \Bar{V} - \rho M_{\sigma(i)} \exp(-\alpha \rho  \Bar{V}  )) -   \frac{\rho  }{M_{\sigma(i)}}\exp(-\alpha \rho \Bar{V}) \\
% & - \frac{1}{M_{\sigma(i+1)}}\exp( - \alpha \rho^2 \Bar{V} - \rho M_{\sigma(i+1)} \exp(-\alpha \rho  \Bar{V}  )) -   \frac{\rho  }{M_{\sigma(i+1)}}\exp(-\alpha \rho \Bar{V}) 
% }
Now, letting $ k = \exp(-\alpha \Bar{V}  ) <1 $ and following the analogous steps from the profit-maximizing case, $g(x) = -\frac{1}{x} + \frac{\exp( - \exp(-\alpha  \Bar{V}  )x ) }{x}$ is increasing in $x$. Thus, the monotonicity holds for $\rho=1$ as in the profit-maximizing case and continuity implies that there exist a $\rho_1$ such that for all $\rho \in (\rho_1, 1]$, the ascending ranking in $(v_i - c_i)$ is optimal.

%\subsection*{Section \ref{sec:implication}} 
\subsubsection*{Proof of Proposition \ref{prop_MonoPrice}}\label{pf:prop_MonoPrice}

We prove Proposition \ref{prop_MonoPrice} by backward-induction. We'll be using Corollary \ref{cor:prop_pp_eqm_exp} that $V^p_i=V^c_i$ in the calculation below. 

 We first prove the monotonicity of the continuation valuations in the customer patience parameter $\rho$. Specifically, we show that under any given ranking $\bsigma$ and in any given stage $i\in\{1,\dots,N-1\}$, the continuation value  $V^{p}_{i+1}$ is monotone increasing in $\rho$.

 \begin{itemize}
    \item \textbf{Inductive hypothesis:} $V^c_i(V^p_i)$ is (weakly) monotone increasing in $\rho$, i.e., $\pap{V^c_i }{\rho} (\pap{V^p_i }{\rho})\geq 0 $ for all $i\in\{N,N-1,\dots,1\}$. 
    
    \item \textbf{Baseline case:} When $i=N$, the expected consumer surplus (which equals to the platform profit by Corollary \ref{cor:prop_pp_eqm_exp}) is given by
    \[ V_N^c = V_N^p = \frac{1}{\alpha} \exp(\alpha (v_N - c_N)-1)  , \]
     which is constant and is weakly increasing in $\rho.$

    \item \textbf{Induction:} In period $i$, $i = N-1, \cdots, 1$, the expected value is
    \eqss{V_i^c = V_i^p &= \frac{1}{\alpha}\PP_i + \rho V_{i+1}^c\\
    & = \frac{1}{\alpha} \exp(-\alpha(2\rho V_{i+1}^c + c_i - v_i +\frac{1}{\alpha})) + \rho V_{i+1}^c, }
    and
    \eqss{   \pap{V_i^c }{\rho}=\pap{V_i^p }{\rho} & = (-2\rho )\exp(-\alpha(2\rho V_{i+1}^c + c_i - v_i +\frac{1}{\alpha})) \pap{V_{i+1}^c }{\rho} +  V_{i+1}^c + \rho \pap{V_{i+1}^c }{\rho}\\
     & = \rho \pap{V_{i+1}^c }{\rho} [1 - \frac{2}{e}\exp(-\alpha(2\rho V_{i+1}^c + c_i - v_i ))] + V_{i+1}^c .
     }
     By the inductive hypothesis, $\pap{V_{i+1}^c }{\rho}$ is non-negative. In addition, $2\rho V_{i+1}^c + c_i - v_i $ is non-negative because $ c_i \geq v_i$ and $2\rho V_{i+1}^c$ is non-negative. Therefore,   $ [1 - \frac{2}{e}\exp(-\alpha(2\rho V_{i+1}^c + c_i - v_i ))]$ is non-negative, which completes the proof.
 \end{itemize}
 Combining this result with corollary \ref{cor:prop_pp_eqm_exp}, we have, $p^*_i=c_i+\frac{1}{\alpha}+\rho V^p_{i+1}$, where $V^p_{i+1}$ is increasing in $\rho$. It follows that each product's optimal price is an increasing function of the consumer patience parameter $\rho$.

\section{Web interface: On the relationship between valuations and prices within a page}  

\label{app_twoproduct}
In this Appendix we analyze the profit-maximizing pricing of two products within a single page ($k=2$) to better understand the surprising inverse relationship between prices and expected valuations within a page. As in Section \ref{sec:implication}, we assume the private valuations are exponentially distributed with unit mean $c_1=c_2=0$. The following Proposition shows that the product with the higher expected valuation is priced \emph{lower}.  \\

\begin{prop}
For a profit-maximizing VA, $v_1 > v_2$ implies $p_1 < p_2$.  
\end{prop}

\textbf{Proof}: The prices $p_1$ and $p_2$ are set to maximize the seller's expected profit $\PP_1p_1+\PP_2p_2,$ $\PP_i$ ($i=1,2$). The probability that the consumer purchases product $i$ is given by 
\eqs{\label{probweb1}\PP_i &= \int^{\infty}_{p_i-v_i}\int^{x+v_i-p_i-v_j+p_j}_{0} \exp(-(x+y))\dd y \dd x  \nonumber\\
&= \int^{\infty}_{p_i-v_i}\exp(-x) (1-\exp(-(x+v_i-p_i-v_j+p_j))\dd x\\
& = \exp(-(p_i-v_i)) - \frac{1}{2} \exp(v_i+v_j-p_i-p_j). \nonumber}    

\noindent The first order conditions simplify to
\eqs{\label{focp1}p_1 &= 1 + \dfrac{p_2}{2\exp(p_2-v_2)-1}, \text{ and}\\ \label{focp2}
p_2 & = 1 + \dfrac{p_1}{2\exp(p_1-v_1)-1}.}
Equations \eqref{focp1}-\eqref{focp2} imply that $p_1$ is increasing in $v_2$ and decreasing in $p_2$, and that $ 1 < p_1, p_2 < 2$. Rearranging \eqref{focp1}-\eqref{focp2}, we have 
\eqs{
\label{eq_foc_web_1}\frac{p_1+p_2-1}{p_1-1} &= 2\exp(p_2-v_2), \text{ and}\\
\label{eq_foc_web_2}\frac{p_1+p_2-1}{p_2-1} &= 2\exp(p_1-v_1).
}
Dividing equation (\ref{eq_foc_web_1}) by equation (\ref{eq_foc_web_2}) and rearranging, we get $\frac{\eta(p_1)}{\eta(p_2)} = \exp(-(v_1-v_2)) <1,$ where $\eta (p) = \frac{p-1}{\exp(p)}$. Since $\eta(p)$ is an increasing function for $p<2$, $\eta(p_1) < \eta (p_2)$ implies $p_1<p_2$, which completes the proof.\\

\begin{figure}[H]
  \centering
  \subfloat{\includegraphics[width=0.6\textwidth]{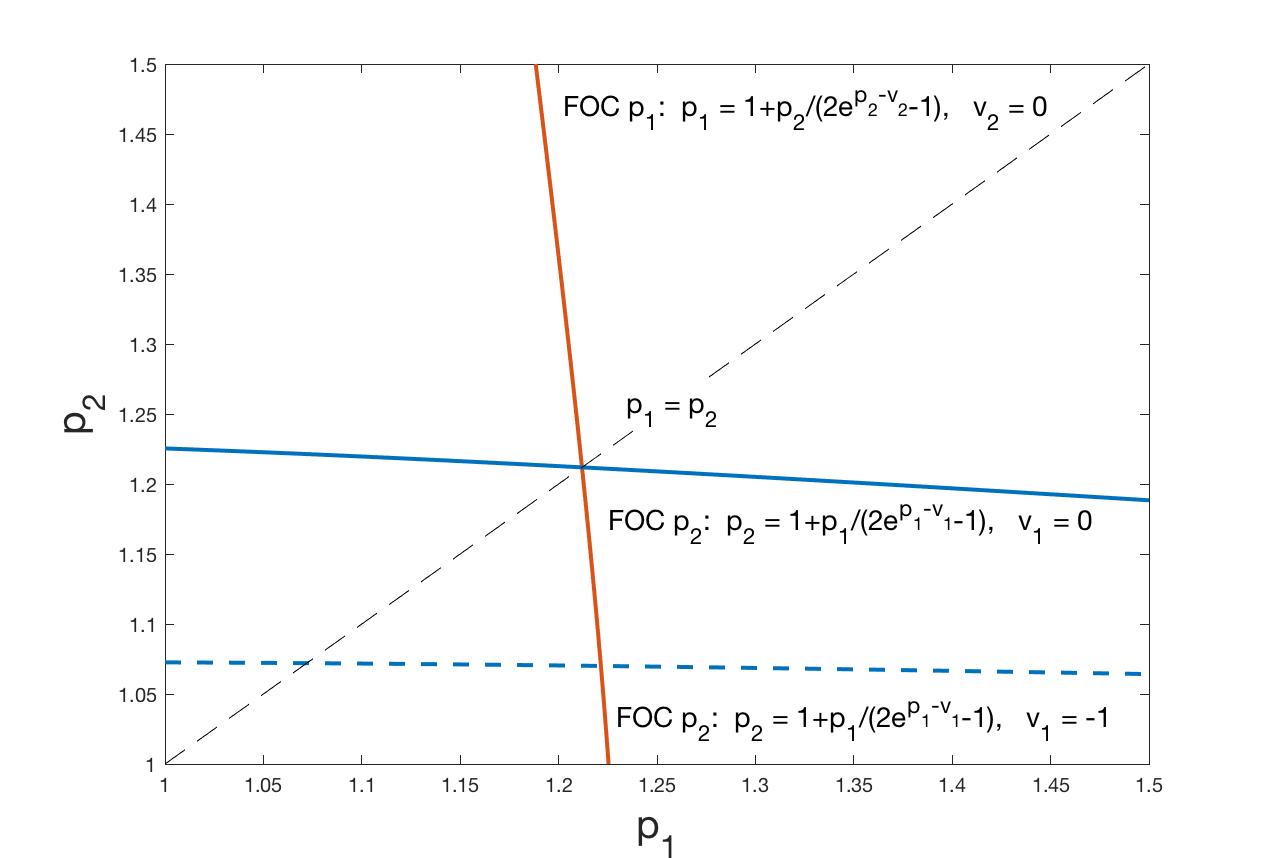}}  \caption{Equation \eqref{focp1}-\eqref{focp2} with different parameters $v_i$. The red and the blue lines correspond to equations \eqref{focp1} and \eqref{focp2}, respectively, when $v_1=v_2=1$. The dotted blue line corresponds to equation \eqref{focp2} with $v_1'>v_1$.}
  \label{web_surplussplit}
  \end{figure}

A more intuitive way of reaching this result is by starting with the case of two products with equal valuations, $v_1=v_2$, and increasing $v_1$.  Obviously, when $v_1=v_2$, $p^*_1=p^*_2$. Now increase $v_1$ to $v'_1>v_1$. First, notice that this has no direct effect on $p_1$: by equation (\ref{eq_foc_web_1}), $p_1$ is not a function of $v_1$. Raising $v_1$ does increase $p_2$: by equation (\ref{eq_foc_web_2}), $p_2$ is an increasing function of $v_1$. Now, increasing $p_2$ leads to a decrease in $p_1$: by equation (4), $p_1$ is a decreasing function of $p_2$. We thus moved from an initial equilibrium with $p_1^*= p_2^*$ to a new equilibrium ($p_1',p_2'$) with $p_1'<p_1^*$ and $p_2'>p_2^*$ (see Figure \ref{web_surplussplit}). Thus, prices have moved in the opposite direction to that of expected valuations.  \\

